\documentclass{aa}
\usepackage{lineno}
\linenumbers

\usepackage{txfonts}
\usepackage{hyperref}
\usepackage{natbib}

\usepackage{multirow}
\usepackage{amsmath}
\usepackage{bm}
\usepackage{enumerate}
\usepackage{booktabs}
\usepackage{orcidlink}

\defcitealias{poggianti+2017b}{P17b}
\defcitealias{peluso+2022}{P22}

\usepackage{subfig}
\usepackage{appendix}
\titlerunning{A\&A,}
\authorrunning {Peluso, G., et al.}
\graphicspath{{Figures/}}

\newcommand{\oiii}{[\ion{O}{iii}]}
\newcommand{\sii}{[\ion{S}{ii}]}

\newcommand{\nii}{[\ion{N}{ii}]}
\newcommand{\oi}{[\ion{O}{i}]}
\newcommand{\oii}{[\ion{O}{ii}]}

\begin{document}



\title{The interplay between Active Galactic Nuclei and Ram-pressure stripping: spatially-resolved gas-phase abundances of stripped and undisturbed galaxies}


   \author{Giorgia Peluso \thanks{\email{giorgia.peluso@inaf.it}}
          \orcidlink{0000-0001-5766-7154}\,
          \inst{1,6}
          \and
          Benedetta Vulcani
          \orcidlink{0000-0003-0980-1499}\,
          \inst{2}
          \and
          Mario Radovich
          \orcidlink{0000-0002-3585-866X}\,
          \inst{2}
          \and
          Alessia Moretti
          \orcidlink{0000-0002-1688-482X}\,
          \inst{2}
          \and
          Bianca M. Poggianti
          \orcidlink{0000-0001-8751-8360}\,
          \inst{2}
          \and
          Peter Watson
          \orcidlink{}\,
          \inst{2}
          \and
          Ayan Acharyya
          \orcidlink{0000-0003-4804-7142}\,
          \inst{2}
          \and
           Augusto E. Lassen
           \orcidlink{0000-0003-3575-8316}\,
          \inst{2}
          \and
          Marco Gullieuszik
          \orcidlink{0000-0002-7296-9780}\,
          \inst{2}
          \and
          Jacopo Fritz
          \orcidlink{0000-0002-7042-196}\,
          \inst{4}
          \and
          Alessandro Ignesti
          \orcidlink{0000-0003-1581-0092}\,
          \inst{2}
          \and
          Neven Tomi\'{c}i\'{c}
          \orcidlink{0000-0002-8238-9210}\,
          \inst{5}
          \and
          Ivan Delvecchio
          \orcidlink{0000-0001-8706-2252}\,
          \inst{6}
          \and
          Amir H. Khoram
          \orcidlink{0009-0009-6563-282X}\,
          \inst{6,7}
          }

   \institute{INAF - Osservatorio Astronomico di Brera, via Brera, 28, 20121, Milano, Italy
    \and
        INAF-Osservatorio Astronomico di Padova, Vicolo Osservatorio 5, 35122 Padova, Italy
    \and
        Dipartimento di Fisica e Astronomia, Universit\`a di Padova, Vicolo Osservatorio 3, 35122 Padova, Italy
    \and 
        Instituto de Radioastronomía y Astrofísica, UNAM, Campus Morelia, A.P. 3-72, C.P. 58089, Mexico
    \and 
        Department of Physics, Faculty of Science, University of Zagreb, Bijenicka 32, 10 000 Zagreb, Croatia 
    \and 
        INAF - Osservatorio di Astrofisica e Scienza dello Spazio Bologna, Via Piero Gobetti, 93/3, 40129 Bologna, Italy
    \and   
        Dipartimento di Fisica e Astronomia "Augusto Righi", Universit\`a di Bologna, via Piero Gobetti 93/2, I-40129 Bologna, Italy 
        }

\abstract
{The gas-phase oxygen abundance of the circumnuclear regions around supermassive black holes (SMBH) has been claimed to be affected by the presence of an Active Galactic Nucleus (AGN). However, an agreement on how this happens has not been reached yet. Some studies measure higher metallicities in the nuclear regions of AGN hosts than in those of star-forming (SF) galaxies, while others find the opposite result.}
{In this work, we explore whether the interplay between AGN activity and the Ram-Pressure Stripping (RPS) acting in the cluster environment can alter the metallicity distributions of nearby ($z < 0.07$) galaxies.}
{We measure the spatially resolved gas-phase oxygen abundances of 10 AGN hosts experiencing  RPS from the GASP survey and 52 AGN hosts located in the field and hence undisturbed by the RP drawn from the MaNGA DR15. To measure the oxygen abundances in SF and AGN-ionized regions, we present a set of strong emission line (SEL) calibrators obtained through an indirect method in which the \oiii/\sii~ and \nii/\sii~ 
 line ratios, observed and predicted from {\sc Cloudy} photoionization models, were matched through the code {\sc NebulaBayes}.}
{We find that the metallicity gradients of RPS and field AGN do not present significant differences within the errors, but that 2 out of the 10 RP-stripped AGNs show lower oxygen abundances at any given radius than the rest of the AGN sample. Overall, this result highlights that the interplay between AGN and RPS seems not to play a major role in shaping the metallicity distributions of stripped galaxies within 1.5 times the galaxy's effective radius ($r < 1.5 \ R_e$), but larger samples are needed to draw more robust conclusions. 
By adding a control sample of SF galaxies, both experiencing RPS and in the field, we find that the AGN hosts are more metal-enriched than SF galaxies at any given radius, but that the steepness of the gradients in the nuclear regions ($r < 0.5 R_e$) is higher in AGN hosts than in SF galaxies. Particularly, AGN hosts show a metal enrichment in the nuclear regions $\approx$ 1.8 - 2.3 times higher than the enhancement in the disk at $r \sim 1.25 R_e$, regardless of the host galaxy's stellar mass. }
{These results favor the hypothesis that the AGN activity is causing metal pollution in the galaxy's nuclear regions.}

\keywords{Galaxy environments --- Active Galactic Nuclei --- Galaxy chemical evolution}

\maketitle

\section{Introduction} \label{sec:intro}


Typically, gas-phase metallicity (or equivalently, abundance) gradients are observed to be negative in spiral galaxies, consistent with an inside-out growth of the spiral discs through the accretion of primordial material with a time scale that is a function of the galactocentric distance \citep[e.g.][]{matteucci+1989, molla+1996}. In this scenario, the gas accumulates and forms the inner disk, where the high density of the gas results in an efficient star formation. The fast gas reprocessing at the galaxy center leads to a population of old, metal-rich stars in an environment of high-metallicity gas. The outer regions are formed later, and therefore these are characterized by younger, metal-poor stars surrounded by less-enriched material \citep[e.g.][]{dave+2011,gibson+2013,prantzos2000,pilkington+2012}.

Deviations from the pure inside-out scenario, thus from the negative slopes, are also commonly observed \citep{dekel+2013,dekel+2014,tissera+2018,grisoni+2018}. 
A decrease or a nearly flat distribution of the abundance in the innermost region of discs was first observed by \cite{belley1992}, while the flattening of the metallicity gradient in the nuclear region of the most massive spiral galaxies was also found in other works using CALIFA \citep[e.g.][]{zinchenko+2016} and MaNGA data \citep[e.g.][]{belfiore+2017,khoram+2024b,kewley+2010}.
Such deviations may originate from the action of other mechanisms, aside from the \emph{in-situ} production through stellar processes, such as loss of metals through outflows in the form of galactic winds \citep{heckman2000,veilleux+2017,tremonti+2007, feruglio+2010}
or transport of pristine gas through gas inflows into the galaxy from the circumgalactic (CGM) and
the intergalactic medium \citep[IGM,][]{prochaska+2017, tumlinson+2017}, or exchange between the halo and the galaxy due to interacting processes such as mergers \citep[e.g.][]{barnes+1991, barnes+1996}
and ram-pressure stripping (RPS) \citep{hughes+2013,franchetto+2020,franchetto+2021,khoram+2024a}.
 Another process that can alter the distribution of metals is the activity of an Active Galactic Nucleus (AGN), even if there are discrepant observations on how this happens. 
 The nuclear regions of AGN hosts have been observed to be either metal-rich \citep[e.g.][]{thomas+2019,peluso+2023} or metal-poor \citep[e.g.][]{storchi-bergmann+1998,armah+2023} with respect to those of SF galaxies.
A possible interpretation to explain a lower metallicity around the nuclear supermassive black hole (SMBH) is the accretion of metal-poor gas that feeds the AGN \citep[][]{storchi-bergmann+2019}. On the other side, the metal enrichment due to AGN could be due to either an in situ top-heavy initial mass function (IMF) in the accretion disk around the SMBH \citep[e.g.][] {nayakshin+2005} or dust destruction in the broad line region (BLR), which releases metals into the ISM \citep{maiolino-mannucci2019}. 
AGN-driven outflows may also contribute by re-distributing the gas from the BLR on parsec (pc) scales \citep[e.g.][]{dodorico+2004} to the Narrow Line Region (NLR).
Indeed, many works have shown that AGN-driven outflows are typically not as effective as expected in expelling metals from the galaxy disk \citep[see also][]{bischetti+2019a, bischetti+2019b,feruglio+2017,veilleux+2017}, and despite its high velocities, most of the gas does not escape the disk \citep{arribas+2014,swinbank+2019}. 
On the other hand, ionized gas outflows detected through optical emission lines are frequently observed in the NLR \citep[e.g.][]{veilleux+1987, unger+1987,pogge+1988}.

Only a few attempts have been made to measure the spatially resolved abundances in AGN host galaxies. \cite{thomas+2018a} studied the spatially resolved abundances in the extended narrow-line regions (ENLR) in four “pure Seyfert” galaxies from the Siding Spring Southern Seyfert Spectroscopic Snapshot Survey \citep[S7,][$z<0.02$, i.e. reaching a spatial resolution up to 408 pc]{dopita+2015}, finding steep inverse or flat metallicity gradients depending on the single-case, indicative of heterogeneity in the mechanisms shaping the observed distributions. 
\cite{amiri+2024} found a positive gradient in the Seyfert NGC 7130 
but with the lowest value in the most central region dominated by the AGN ionization. MUSE data were able to resolve the NLR due to the vicinity of the source ($z = 0.016$) which allowed to achieve a spatial resolution of $\sim$ 0.3 kpc.  
Going beyond a single case study and exploiting larger samples, \cite{donascimento+2022} measured the abundance gradients in a sample of Seyfert from the inner AGN-dominated regions up to the outer star-forming (SF) regions, exploring integral-field spectroscopy (IFS) data from the MaNGA DR15 survey \citep[][$z < 0.15$]{bundy+2015}. These maps were characterized by a spatial resolution of $1-4$ kpc. They found differences of $0.16$ to $0.30$ dex between the abundances in the nucleus and the {\sc Hii} regions of the galactic disk. 

In this work, we isolate the AGN hosts in clusters and in the field to explore the effect of the environment on the metallicity gradients.
So far, the role of the RPS in driving the evolution of galaxies, including its AGN activity, has been investigated by a large number of both observational and theoretical studies. Many observational works agree that the RPS initially enhances the SFR in the surviving disk \citep{gavazzi+1995,gavazzi+2001,consolandi+2017,vulcani+2018,vulcani+2020b}. It has been argued that the increased SF may result from an increased density linked to the compression of the gas in the leading side of the disk \citep{tomicic+2018}. \cite{vulcani+2018} measured the SFR-$M_*$ of 42 RP-stripped galaxies, finding a systematic average enhancement (0.2 dex) at any given mass with respect to a control sample of 32 field and undisturbed cluster galaxies. 
These galaxies were drawn from the GASP \citep[GAs Stripping Phenomena in galaxies with MUSE,][]{poggianti+2017a} program, which is a MUSE large program aimed at studying the gas removal processes in massive ($\log M_*/M_\odot \geq 9.0$) galaxies in different environments.
Other than an increased SFR, the GASP sample also showed an enhanced AGN fraction ($\sim$ 46\%) with respect to field galaxies ($\sim$ 28\%) \citep[][]{poggianti+2017a,peluso+2022}, indicative of a starburst-AGN activity connection \citep{matsuoka+2013} in these objects. The galaxies used to compute the above fractions span the mass range $\log (M_*/M_\odot) \geq 10.0$, typical of AGN hosts \citep[e.g.][]{kauffmann+2003}. 
The AGN fraction increases as a function of the tail's length, as indeed the chance for a jellyfish to host an AGN is 56\%, as opposed to the 33\% in galaxies at earlier or later stages of the stripping \citep{peluso+2022}.
To explain this result, both hydrodynamical and cosmological simulations have found growing evidence of angular moment loss induced by the ICM-ISM interaction, which pushes inflows of gas towards the galaxy centers \citep{akerman+2023,schulz+2001,tonnesen+2009,ramos-martinez+2018}, and may light up or enhance the AGN activity.
In particular,
\cite{ricarte+2020} find that the black hole accretion rate  (BHAR) increases strongly, and corresponds with the SFR's peak, when the RPS intensity reaches its maximum. These results suggest a possible link between the increased star formation and AGN activity observed in the GASP sample, which is composed of RP-stripped galaxies hosted by clusters spanning a wide range of velocity dispersions, from Fornax to Coma-like (Poggianti et al. submitted).

However, it is worth mentioning that other works, focusing on single clusters, find contrasting results, hampering the ability to draw more robust conclusions. \cite{roman-oliveira+2019} have shown that only 5 over 70 jellyfish with stellar mass $\log M_*/M_\odot \geq 9.0$ seem to show AGN activity in the A901/2 cluster at $z=0.165$ \citep[see also][]{boselli+2022}.
However, as also shown by simulations, the period of activity of the SMBH is expected to be relatively short in the whole life cycle of a galaxy undergoing RPS \citep{ricarte+2020}. Indeed, after the intensity peak of the RPS, both SF and the BHAR decrease rapidly. 

To explore the metal distribution in the case of an AGN-RPS interplay as described above, spatially-resolved abundance gradients of a large sample of RP-stripped galaxies are needed, as well as gradients from a control sample of non-stripped AGN hosts to perform a comparison with the RPS sample. 
The use of IFS data allows us to retrieve the metallicity gradients necessary to investigate in detail the metal production and pollution in these objects, and to disentangle ionization from the AGN or the stars. 
Optical spectroscopy, in particular, is needed to perform measurements of the oxygen abundance, which in general happens through a direct method  (also called $T_e$-method) involving auroral lines \citep[such as \oiii$\lambda$4363, see e.g.,][]{dors+2020b} or an indirect method involving photoionization models \citep[see e.g.][]{thomas+2019}. 
The direct $T_{\rm e}$-method has been widely adopted to measure abundances in {\sc Hii} and AGN-ionized regions \citep[e.g.,][]{dors+2020a}.
However, although AGNs have a high-ionization degree, their high \citep[e.g.,][]{groves+2006} metallicity leads to faint auroral lines (such as the  \oiii~$\lambda$4363), hampering the use of the $T_{\rm e}$-method. Some works \citep[e.g.,][]{thomas+2019,perez-diaz+2021} have developed AGN and stellar photoionization models using the same method, and compared observed and predicted line ratios by making use of the Bayesian inference with codes such as {\sc Nebulabayes} \citep{thomas+2018a} or \textsc{{\sc Hii}-Chi-Mistry} \citep{perez-montero+2014,perez-montero+2019}.  Strong emission-line (SEL) calibrators for the NLR have been derived through photoionization models \citep[e.g.,][]{carvalho+2020,storchi-bergmann+1998} or the direct method \citep{flury-Moran+2020, dors+2021}. However, SEL calibrators obtained by measuring the abundances with the same method and using the same set of measurements in case of ionization from star formation and AGN are still not available in the literature.

In this paper, we present a set of SEL calibrators to measure the oxygen abundance in {\sc Hii} (or equivalently, SF), AGN, and AGN+{\sc Hii} regions classified by the Baldwin, Philiphs, and Telervich \citep[BPT,][]{baldwin+1981} diagram involving the \nii/H$\alpha$ and \oiii/H$\beta$ ratios, also called \nii-BPT.
Exploiting these calibrators, we analyze the spatially-resolved oxygen abundances of stripped and undisturbed galaxies using IFS data from the GASP and MaNGA \citep[Mapping Nearby Galaxies at Apache Point Observatory][]{bundy+2015} surveys.
Our sample comprises galaxies with and without AGN activity, as evident from the BPT classification. We isolate stripped AGN and non-stripped AGN to explore whether the interplay between RPS and AGN activity affects the metallicity distributions. 
In this study, which follows up on our previous work \citep{peluso+2023} where we observed enhanced nuclear metallicity in AGN hosts compared to SF galaxies, we investigate the metal abundances at various galactocentric distances within both AGN and SF galaxies. Our goal is to determine whether the increased metallicity in AGN hosts is limited to the nuclear regions, indicating that it results directly from AGN activity, or if it extends throughout the entire galaxy disks. If the latter is true, it would indicate that the observed enhancement in metallicity may be influenced by other factors, such as the galaxy's star formation history (SFH).

Sections are structured as follows: the observational dataset used for the analysis is described in Section \ref{sec:data}, followed by Section \ref{sec:method} in which the SEL calibrators are presented, Section \ref{sec:result} shows the main results, in Section \ref{sec:discussion} we discuss possible secondary dependences of the results linked to the method used to measure the metallicity or the galaxy's properties characterizing the sample; finally in Section \ref{sec:summary} we summarize and conclude. 

We adopt a \cite{chabrier+2003} initial mass function in the mass range 0.1-100 M$_\odot$. We assume a standard $\Lambda$CDM cosmology with $\Omega_m$ =0.3, $\Omega_{\Lambda}$ = 0.7 and $H_0$ = 70 km $\rm s^{-1}$ Mpc$^{-1}$. Throughout this work, we adopt 12 + $\log$(O/H) = 8.69 as the solar oxygen abundance \citep{asplund+2009}.

\section{Datasets and Galaxy samples} \label{sec:data}


As in \cite{peluso+2023} (P23 hereafter), this paper is based on galaxies selected from the GASP \citep{poggianti+2017a} and MaNGA \citep{bundy+2015} surveys. The former sample allows us to characterize galaxies located in clusters and experiencing RPS, and the latter to define a field sample (FS) to study galaxies  not disturbed by environmental effects.
In this way, we can understand the influence of the RPS phenomenon on the galaxy's properties. Since we are also interested in focusing on the role of AGN activity in regulating the metal content in galaxies, and more in detail in the study of the interplay between RPS and AGN activity, in both samples we separate galaxies hosting AGN activity from those dominated by star formation within the nuclear regions.
In both samples, we select galaxies with a high specific star-formation rate (sSFR $ > 10^{-11} M_\odot$/yr) and with a late-type morphology according to their optical emission.
This first selection leads us to four samples: AGN-RPS, AGN-FS, SF-RPS, and SF-FS, described in detail below.

\subsection{RPS sample}

The RP-stripped galaxies are selected from GASP. The program targeted
114 galaxies in the redshift range 0.04 $<$ z $<$ 0.07 and mostly located in clusters.
Observations were carried out with the MUSE spectrograph, which covers the wavelength range 4650 to 9300 \AA~ and have a seeing-limited spatial resolution of $\sim$ 1 kpc.
The data, reduced with the MUSE pipeline \citep{bacon+2010}, were first corrected for extinction caused by our own Galaxy, estimated at the position of each 
galaxy \citep{schlafly_finkbeiner2011} using the extinction law from \cite{cardelli+1989}. Next, the emission lines were fitted with the code KUBEVIZ \citep{fossati+2016} from the continuum-subtracted and extinction-corrected MUSE spectrum,  as described in \cite{poggianti+2017a}.
Finally, the flux measurements were further corrected for extinction by dust internal to the host galaxy. The correction is derived from the Balmer decrement at each spatial element location adopting the  \cite{cardelli+1989} extinction law and assuming an intrinsic Balmer decrement $\rm I(H\alpha)/I(H\beta) = 2.86$, appropriate for an electron density $n_{\rm e}$=100~cm$^{-3}$ and electron temperature $T_{\rm e} =10^4~K$ \citep{Osterbrock2006}.
Integrated galaxy global properties such as inclination, position angle, and effective radius (${R_{\rm e}}$) are taken from \cite{franchetto+2020}, who measured these quantities by means of I-band MUSE photometry. In particular, the effective radius $R_{\rm e}$ was computed from the luminosity growth curve $L(R)$ of the galaxies, obtained by trapezoidal integration of their surface brightness profiles. 
Stellar masses, which range from $\log M_*/M_\odot = 9.0$ to $\log M_*/M_\odot = 11.5$, are taken from \cite{vulcani+2018} and are computed assuming a \cite{chabrier+2003} IMF using the SINOPSIS spectrophotometric code \citep{fritz+2017}. These quantities are obtained by summing the values of all of the spaxels belonging  to the galaxy disk, as defined in \cite{gullieuszik+2020}.
To quantify the effect of the stripping, we select galaxies at least in an early stage of the stripping, i.e. JType $\geq$ 0.5, where JType is a number ranging from 0 to 4 quantifying the visual evidence for stripping signatures (see Poggianti et al. submitted for further details on the definition of JType).

We select 51 galaxies located in clusters that show signs of stripping.  11/51 hosts an AGN according to the spatially-resolved BPT classification with the \nii /H$\alpha$ versus \oiii /H$\beta$  diagnostic \citep[\nii-BPT,][]{baldwin+1981}  \citep[i.e. AGN-RPS, see also][]{peluso+2022}. JO36 is an obscured AGN detected by \emph{Chandra} as a point-like X-ray source \citep{fritz+2017}, but not directly identified using BPT diagrams due to strong dust absorption and for this reason was not included in the AGN-RPS sample.
The remaining 39/51 galaxies are dominated by SF emission. Since \cite{franchetto+2020} could not retrieve the structural parameters for 4 out of the 39 from their I-band photometry, we end up with an SF-RPS sample of 35 galaxies with $\log (M_*/M_\odot) \geq 9.0$.
All the AGN hosts have stellar masses $\log (M_*/M_\odot) \geq 10.5$, while only 6/35 SF galaxies are in this mass regime.

\subsection{Field sample}
We select galaxies located in the field from the MaNGA \citep[Mapping Nearby Galaxies at Apache Point Observatory][]{bundy+2015} survey, which is an integral-field spectroscopic survey using the BOSS Spectrograph \citep{smee+2013} mounted at the 2.5 m Sloan Digital Sky Survey (SDSS) telescope \citep{gunn+2006}, and in particular we exploited the MaNGA Data Release 15 \citep[DR15][]{bundy+2015}. 
The spectral coverage extends from 3600 to 10300 \AA.

To ensure that galaxies are not affected by RPS, we consider only galaxies located in halo masses lower than M$_{halo} < 10^{13}$ \ {\rm M}$_\odot$, according to the \cite{tempel+2014} environmental catalog. 
We use the online tool MARVIN\footnote{\url{https://www.sdss.org/dr16/manga/marvin/}} \citep{cherinka+2019} to download the elliptical radius and azimuth (\emph{spx\_ellcoo}), and the emission line fluxes (\emph{gflux}). The de-projected coordinates are computed using the ellipticity ($\epsilon$ = 1-b/a) and position angle ($\theta$) measured from the r-band surface brightness. 
The same emission line fluxes are drawn from the drpall-v2\_4\_3 and have S/N $>$ 1.5, which is the value typically adopted in MaNGA \citep{belfiore+2019}. 
Emission lines are fitted with a Gaussian function and are corrected for stellar absorption since the Data Analysis Pipeline \citep[DAP;][]{westfall+2019,belfiore+2019} simultaneously fits the continuum and emission lines with the latest version of the pPXF software package \citep{cappellari+2017}. 
All lines are also corrected for Galactic extinction, using the \cite{schlegel+1998} maps \citep{westfall+2019} and the reddening law of \cite{o'donnell+1994}.  Following the same approach used in GASP, we correct the emission lines for host galaxy dust attenuation using the \cite{cardelli+1989} law.
We discard  spaxels which have a S/N $<$ 3 for any of the emission lines H$\beta$, \oiii~ $\lambda$5007, H$\alpha$, \nii~ $\lambda$6584, \sii~ $\lambda$6716 and \sii~ $\lambda$6731.
We select 429 galaxies located at redshift below z $<$ 0.04, which corresponds to a spatial resolution of $\sim$ 1 kpc inside the MaNGA PSF of 2.5$^{\prime \prime}$ \citep{law+2016}. These are part of the Primary+Color-Enhanced (or Primary+) sample \citep{bundy+2015}, and thus posses a smooth distribution in redshift \cite[see Figure 8 in][]{bundy+2015},
and are also part of the \citet{pace+2019a,pace+2019b} catalog, from which we extract an estimate of the total stellar mass. \citet{pace+2019a,pace+2019b}  make use of the Principal Component Analysis (PCA) to compute the stellar mass within the FoV of MaNGA, which covers the galaxies of the Primary+ sample up to 1.5 effective radius ($R_e$), and the color-mass relation by \cite{bell_dejong+2001} to compute the stellar mass outside the MaNGA FoV, thanks to the NASA Sloan Atlas \citep[NASA][]{blanton+2011} colors and magnitudes.
As already discussed in P23, we checked that in this way stellar masses from MaNGA and GASP are estimated similarly and coherently, and are not affected by systematic effects (e.g., different data reduction and instruments) caused by the use of two surveys.
The final sample spans the stellar mass range 9.0$\leq \log ({\rm M}_* /{\rm M}_\odot)  \leq$11.3. Among the 429 galaxies, 52 are Seyfert/LINER (i.e., AGN-FS) according to the \nii-BPT classification. 377 galaxies are instead classified as star-forming and constitute to the SF-FS (see also P23). AGN hosts have stellar masses $\log (M_*/M_\odot) \geq 10.5$, while 78 out of the 377 SF galaxies exceed this threshold.

\section{Method} \label{sec:method}

\subsection{Photoionization models and the code {\sc Nebulabayes}} \label{sec:models}

We use an indirect method to compute the gas-phase oxygen abundances.
As described in details in P23, we generate photoionization models with the code {\sc Cloudy} v17.02 \citep{ferland+2017} assuming the gas to be ionized by radiation coming from stars ({\sc Hii} models, hereafter) or AGN (AGN models, hereafter), available in the online material. 
We assume a gas density of ${\rm n}_H = 100 \ {\rm cm}^{-3}$,
stellar age $t_* =$ 4 Myr, and a simple power law to model the AGN continuum, with a slope from the infrared to X-ray wavelengths equal to $\alpha$ = - 2.0.
The free parameters are the ionization parameter ($\log U$) and the gas-phase metallicity ($\log Z$, or equivalently, oxygen abundance 12 + logO/H).

To compute the metallicity in Composite (AGN+SF) regions, we mixed the {\sc Hii} and AGN models by means of the expression:
\\
\begin{equation}
f_{AGN} = \frac{R_{AGN}}{R_{{\rm HII}}+R_{AGN}}
\end{equation}


where $R$ is the flux of the reference line, i.e. H$\beta$, thus $R_{AGN}$ is the H$\beta$ flux that arises from the AGN and $R_{{\rm H II}}$ is the H$\beta$ flux that arises from the {\sc Hii} regions. In other words, the parameter ‘AGN fraction’ (f$_{AGN}$) is driving the mixing and this is defined as the fraction of H$\beta$ flux arising from the AGN with respect to the total ({\sc Hii} regions+AGN) H$\beta$ flux. 

For further details on the {\sc Hii}, Composite, and AGN models we refer the reader to Sections 3.1, 3.2, and 3.3 in P23.

\smallskip
\begin{figure*}
    \centering
    \makebox[\textwidth]{
    \includegraphics[scale=0.32]{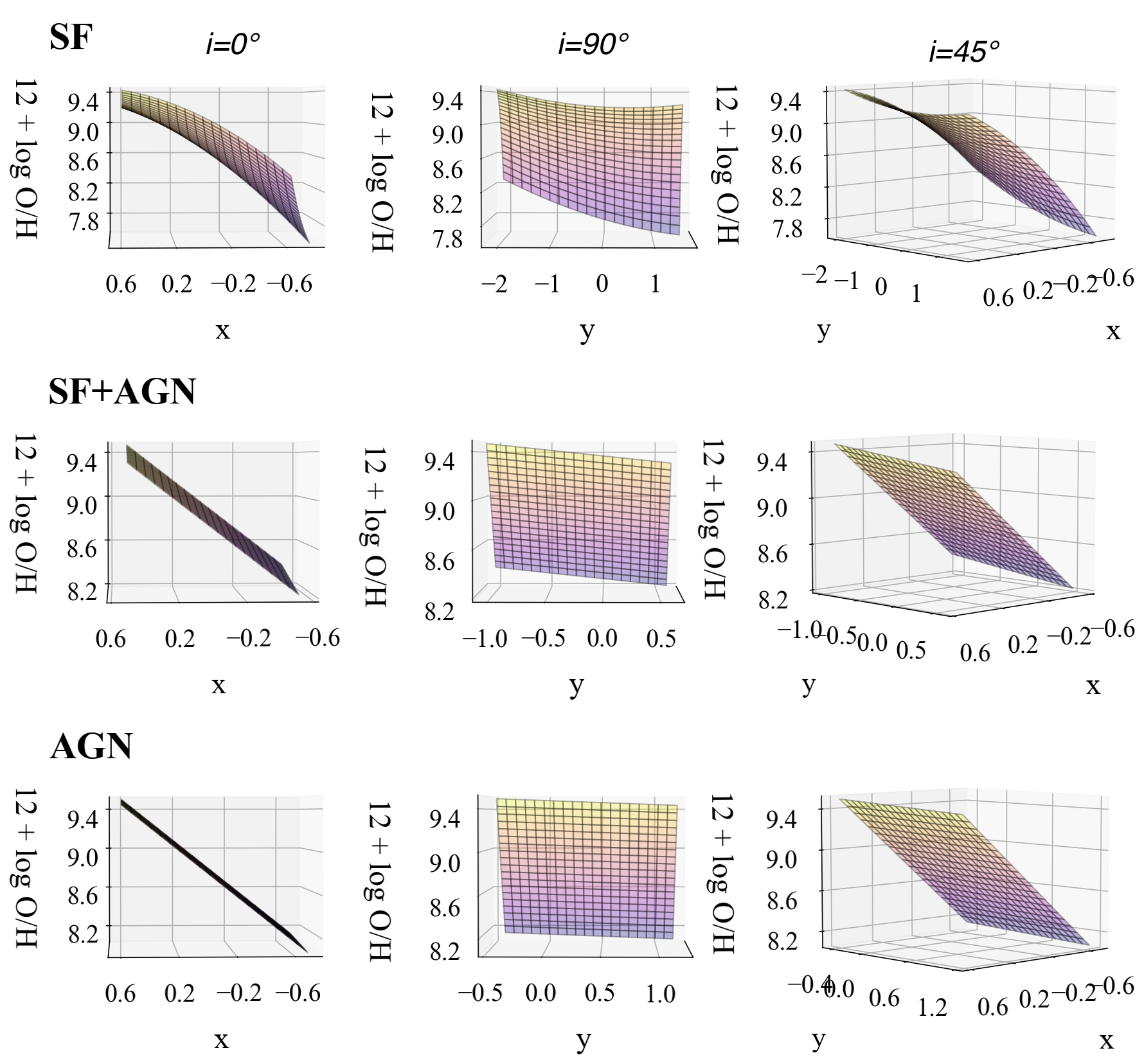}}%
\caption{Surfaces of 12 + log O/H versus $x \equiv \log$(\nii/\sii) and $y \equiv \log$\oiii/\sii, as described by the {\sc Sf} calibrator or equation \ref{eq:sf_calib} (top panels), {\sc Comp} calibrator or equation \ref{eq:comp_calib} (center panels) and {\sc Agn} calibrator or equation \ref{eq:agn_calib} (bottom panels), rotated of an angle of 0$^{\circ}$ (right), 90$^{\circ}$ (center) and 45$^{\circ}$ (left). 
    The surfaces are obtained by fitting with a least-squared method the observed \oiii/\sii~ and \nii/\sii~ line ratios and the 12 + log O/H computed with {\sc Nebulabayes} inside the SF spaxels of all the samples, and the Composite and AGN spaxels in the AGN-RPS and AGN-FS. 
    Cells are color-coded according to the value of 12 + log O/H.} 
    \label{fig:3d-plots}
\end{figure*}

Then, we use the Bayesian code {\sc Nebulabayes} \citep{thomas+2018a, thomas+2018b, thomas+2019} to find the best model that reproduces the observed emission. More specifically, the code takes as input a set of observed line ratios, and corresponding uncertainties, to return the parameters of photoionization models (such as metallicity, ionization parameter, gas density, etc.) that best match the observed emission. 
To do so, the code computes
the posterior probability distribution function (PDF) by multiplying the so-called likelihood with a prior. The likelihood is the probability that a particular set of model parameter values are truly representative of the observed line ratios \citep[for further details on the Bayesian calculation, see][]{thomas+2018a}.
The prior, which is a function assigned by hand, was chosen to be uniform, since the combination of lines used in input to the code is already sufficient to give a constraint on the free parameters.
Finally, the `best model’ is the point in the parameter space which maximizes the posterior PDF.

As input, we give the observed lines \oiii$\lambda$5007 (hereafter \oiii) and \nii$\lambda$6584 (hereafter \nii) divided by a reference line, which in our case was the sulfur duplet \sii$\lambda\lambda$(6716+6731) (\sii~ hereafter). This essentially translates into comparing predicted and observed \oiii/\sii and \nii/\sii ratios (see also P23).
As shown in Figure 1 of P23, these two-line ratios, coupled together, work excellently in breaking any degeneracy between $\log U$ and $\log Z$, for a fixed gas density (e.g. $n_{H} = 100 \ \rm cm^{-3}$). 

\subsection{Metallicity calibrators in SF and AGN-ionized regions} \label{sec:calibrators}

\begin{figure*}
    \centering
    \makebox[\textwidth]{
    \includegraphics[scale=0.25]{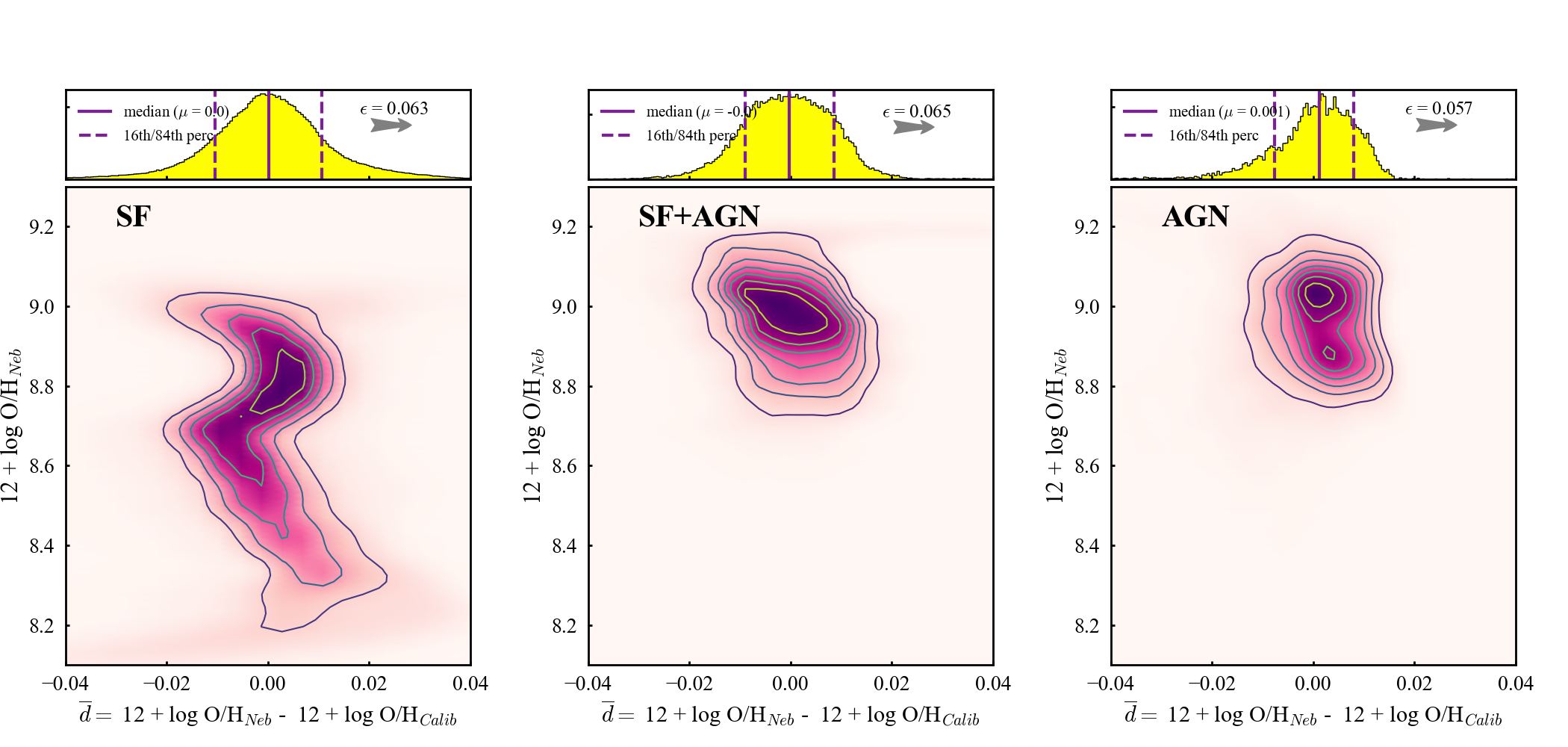}}%
    \caption{12 + log O/H computed by {\sc Nebulabayes} (12 + log O/H $_{Neb}$) versus the difference $\overline{d}$ between 12 + log O/H $_{Neb}$ and the metallicities computed through the (\emph{from left to right})  SF, COMP (AGN+SF), and AGN calibrators  (12 + log O/H $_{Cal}$).
    These distributions have been smoothed by a 2D Gaussian Kernel Density Estimate (KDE), where the black contours outline density curves and darker shades of pink indicate higher densities.
    Yellow histograms on the top panel showcase the distribution of $\overline{d}$ along with the 16th/84th percentiles (purple dotted lines) of the distributions.
    Values of 12 + log O/H $_{Neb}$ and 12 + log O/H $_{Cal}$ are consistent within the errors. 
    Gray arrows point toward the value of the median error on 12 + log O/H $_{Neb}$ ($\epsilon$ in top panels), which is outside the range of values shown in the x-axis. This indicates that $\epsilon$ is 3$\times$ higher than the maximum difference between 12 + log O/H $_{Cal}$ and  12 + log O/H $_{Neb}$.}
    \label{fig:errors}
\end{figure*}

Despite the fact that the {\sc Nebulabayes} analysis represents an excellent technique to measure metallicity robustly, this routine is relatively expensive in terms of time and computational power. 
Because of this, we derive three simple equations, one to derive the oxygen abundance in regions ionized by stars, ({\sc Sf} calibrator), one in the AGN-ionized regions ({\sc Agn} calibrator), and one in the Composite regions ({\sc Comp} calibrator). 
Computing the abundances with these equations, which are a linear combination of the observed \oiii/\sii~ and \nii/\sii~ line ratios, reduces the computational time from hours to minutes.
We obtain them by fitting the relation between 12 + log O/H as computed by {\sc Nebulabayes} and the \oiii/\sii~ and \nii/\sii~ line ratios observed in {\sc Hii}, AGN and {\sc Hii}+AGN regions (in the \nii-BPT).
To obtain the {\sc Comp} calibrator, we consider only line ratios inside 'Composite' spaxels in which the photoionization models predict the parameter $f_{AGN}$ to be 0.2 because the majority (85 \%) of the 'Composite' spaxels have this value of $f_{AGN}$.

The 3D surfaces  (shown in Figure \ref{fig:3d-plots}) are fitted with a least-square method and can be described by the following equations:

\begin{center} 
12+log \ O/H$_{\rm P24, SF} = $ \\
8.78 + 0.97x  -  0.11y - 0.39x$^2$ + 0.09xy + 0.02y$^2$
\end{center}
\begin{equation} \label{eq:sf_calib}
\\
\\
\end{equation}

\begin{center}
12+log \ O/H$_{\rm P24, AGN}$ = 
     8.85 + 1.06x - 0.04y
\end{center}
\begin{equation} \label{eq:agn_calib}
\end{equation} 
 \begin{center}
 12+log \ O/H$_{\rm P24, COMP}$ = 
 8.83 + 1.07x  + 0.10y
\end{center}
\begin{equation} \label{eq:comp_calib}
 \end{equation}



where $x\equiv\log$ (\nii/\sii) and $y \equiv\log$ (\oiii/\sii).
The linear correlation in the 3D plane between the line ratios and the oxygen abundance is applicable up to 12 + log O/H = 8.2, while disappears for lower metallicities. Thus, the validity range is set to be 12 + log O/H $>$ 8.2.
 These calibrators (hereafter P24 calibrators) are both easier to apply and considerably reduce the computational time needed to obtain spatially resolved maps. From now on, we will use the values obtained by Eq 1, 2, 3, to describe metallicities in the various regions.



Figure \ref{fig:errors} demonstrates that the surfaces of equations \ref{eq:sf_calib}, \ref{eq:agn_calib} and \ref{eq:comp_calib} successfully reproduce the spaxel-by-spaxel predictions by photoionization models. The distribution of the difference $\overline{d}$ between the abundances computed with the calibrators (12 + log O/H$_{Calib}$) and with photo-ionization models coupled with {\sc NebulaBayes} (12 + log O/H$_{Neb}$) is plotted versus the values of 12 + log O/H$_{Neb}$. 
Depending on the value of 12 + log O/H$_{Neb}$, $\overline{d}$ peaks between 0.02 dex and -0.02 dex in the SF spaxels, while the median $\mu$ is constantly $\sim$ 0.0 dex at any 12 + log O/H$_{Neb}$ in the AGN and Composite spaxels. The yellow histograms in the top panels show that the median value of $\overline{d}$ ($\mu$) (over all the values of 12 + log O/H$_{Neb}$) is 0.0 in all three panels, with a scatter of $\pm$ 0.01,  $\pm$ 0.009 and $\pm$ 0.008 dex given by the 25th/75th percentiles for the SF, AGN+SF, and AGN spaxels.\footnote{To further test the robustness of our results, we ensured that the metallicity gradients and the derived relations presented throughout the paper are perfectly consistent when computing the metallicity by means of the calibrators or the photoionization models coupled with NebulaBayes. We remind the reader that the main assumptions on the models are the stellar age of $t_* = 4 \times 10^6$ yrs, the gas density $n_{H} = 10^2$ cm$^{-3}$ and a slope of the AGN continuum $\alpha = -2.0$.}
Gray arrows point out that the error computed by {\sc Nebulabayes} for the metallicity estimates ($\epsilon$, which is the maximum between the 16th and 84th percentiles of the posterior PDF) is a factor $\sim$ 3 times higher than the maximum difference.
Thus, the uncertainty on the metallicity estimates remains the one associated to the photo-ionization models, as indeed the error introduced by the linear fitting is negligible.
Only one galaxy, JW100, part of the AGN-RPS sample, shows significant deviations from the calibrators' planes ($\mu >$ 0.1 dex).  This galaxy is also notably characterized by peculiar ISM conditions, as deduced by the high X-ray luminosity in the tail \citep[e.g. ten times higher than other tails in RP stripped galaxies, see][]{sun+2021,poggianti+2019}, and the still-unclear excess of \oi$\lambda$6300 emission with respect to  what is typically observed in extra-nuclear {\sc Hii} regions \citep{poggianti+2019}. Because of these reasons, the galaxy is from now on discarded from the AGN-RPS sample.




\subsection{Metallicity Gradients and Slopes}

Spatially-resolved maps (with a resolution of $\sim$ 1 kpc) of the oxygen abundance are obtained by making use of the {\sc Sf}, {\sc Comp}, and {\sc Agn} calibrators (presented in Section \ref{sec:calibrators}) applied to spaxels classified accordingly to the \nii-BPT.




Once the metallicities are computed for any given spaxel, we divide each galaxy into concentric annuli with width 0.3 $R_e$, except for the first central annulus having an outer radius of $r = 0.5 \ R_e$. 
We chose this aperture as this is always larger than the PSF (e.g., aperture with diameter d $\sim$ 2.5\arcsec~ in MaNGA, and d $\sim$ 1\arcsec in GASP), and therefore includes a well-resolved galactic region.
To obtain the metallicity gradients, we compute  the median value of 12 + log O/H of all the spaxels (AGN/Composite/SF) inside each annulus and the corresponding 25th/75th percentiles of the 12 + log O/H distribution. We set a threshold on the minimum number of spaxels used to compute the medians and errors, which is $N_{{\rm th}}$ = 10 spaxels.
Finally, we plot the median values of 12 + log O/H versus the mean value of $R/R_e$ inside a given annulus.
The outermost radius ($r_{out}$) of the last annulus is set to be $r \sim 2.5 \ R_e$. However, in some cases the MaNGA data 
cover the  galaxy disks only up to 
a radius $R \sim 1.5 \ R_e$ or  $R \sim 2.2 \ R_e$ \citep[see][for details]{bundy+2015}.
For this reason, to consistently compare the results among all the galaxies of the different samples, we will consider abundances up to $r_{out} \sim$ 1.5 $R_e$,  the maximum distance covered in all galaxies.

To quantify the abundance variation throughout the galaxy, we define the slope $\Delta \alpha$ of the radial profile as:
\begin{equation} \label{eq:slope}
\Delta \alpha = \delta (O/H) / \delta R
\end{equation}
with $\delta \ (O/H) = 12 + \log \ O/H_{r = 1 R_e} - 12 + \log \ O/H_{r < 0.5 R_e}$  and  $\delta R$ = 1 $R_e$. We compute the slopes only within the galaxy's optical radius ($R < 1 \ R_e$). In this way, we expect not to be affected by the flattening typically observed at very large galactocentric radii \citep[$R>2 R_e$), suggestive that galactic outskirts have accreted pre-enriched material, see][]{maiolino-mannucci2019, franchetto+2021}.






\section{RESULTS} \label{sec:result}

\begin{figure*}
    \makebox[\textwidth]{
    \includegraphics[scale=0.32]{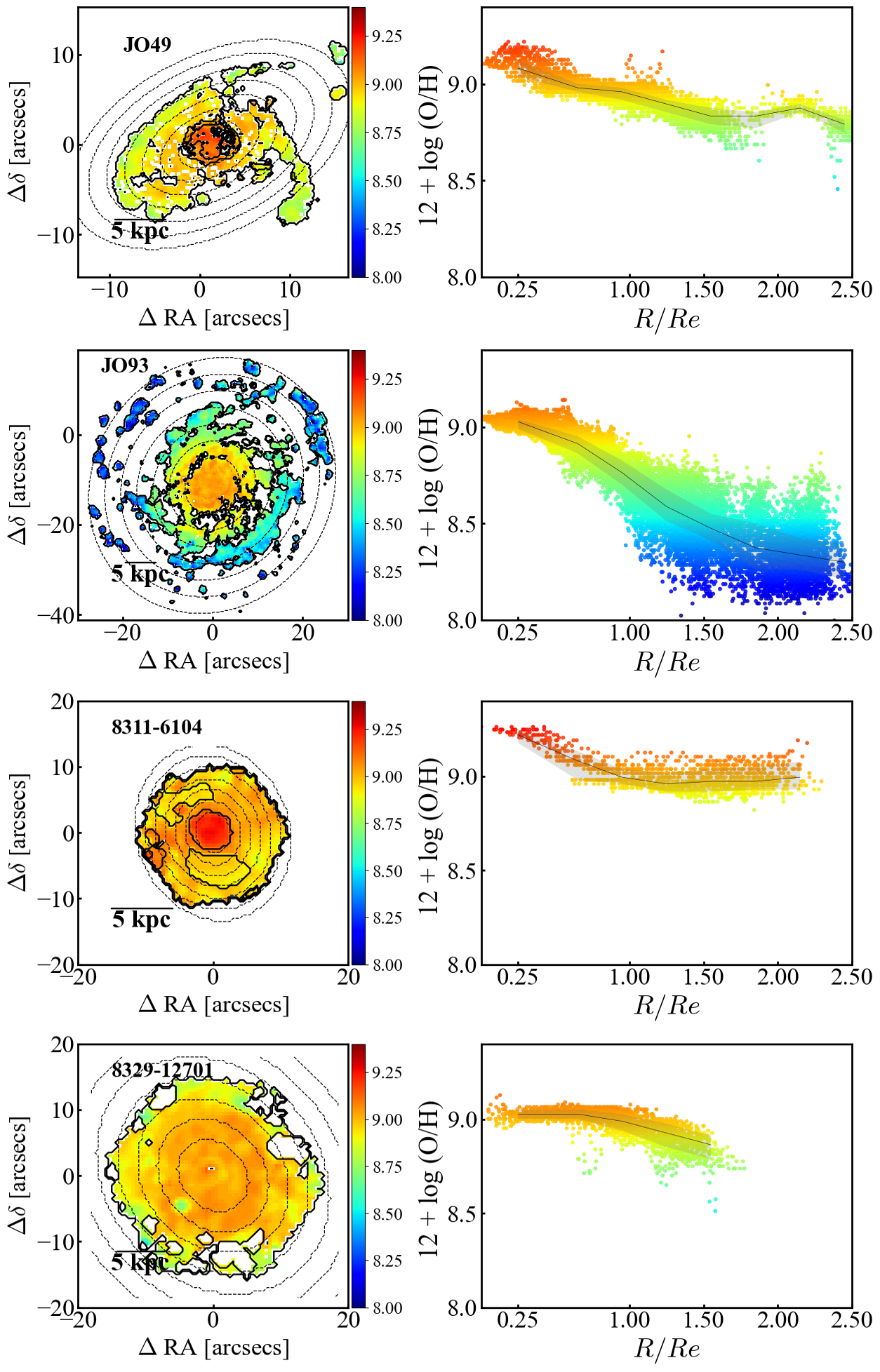}}%
    \caption{\emph{From top to bottom}: metallicity maps (left panels) and metallicity gradients (right panels) of the cluster-AGN host galaxy (JO49, $\log M_*$= 10.68, $z=0.0451$), cluster-SF galaxy (JO93, $\log M_*$ = 10.54, $z=0.037$), field-AGN galaxy (’8311-6104’, $\log M_*$ = 10.67, $z=0.027$) and the field-SF galaxy ('8329-12701', $\log M_*$ = 10.99, $z = 0.035$). On the left, the black contours are overplotted on the metallicity map to divide SF, Composite, and AGN-like regions, as classified by the \nii - BPT. The gray dotted ellipses are the annuli that cover the galaxy up to $R_e \sim$ 2.5, proceeding with a step of 0.3 dex, except for the central annulus which has an inner radius of $r_{in}$ = 0 $R_e$ and an outer radius of $r_{out}$ = 0.5 $R_e$. 
    Inside each annulus, the median value of the 12 + log O/H distribution is computed, along with the 25th/75th percentiles, considering all the 12 + log O/H  values inside the SF/Composite/AGN spaxels. On the right, the metallicity profile (gray line) is obtained by joining the median values of 12 + log O/H of each annulus, plotted versus the mean value of R/R$_e$ inside a given annulus, and the shaded area which covers the range of 12 + log O/H values within the 25th/75th percentiles. In both panels, the color coding is set according to the value of 12 + log O/H, ranging from 8.0 to 9.4.}
    \label{fig:gasp+manga}
\end{figure*}


\begin{figure*}
     \centering
     \makebox[\textwidth]{
    \includegraphics[scale=0.47]{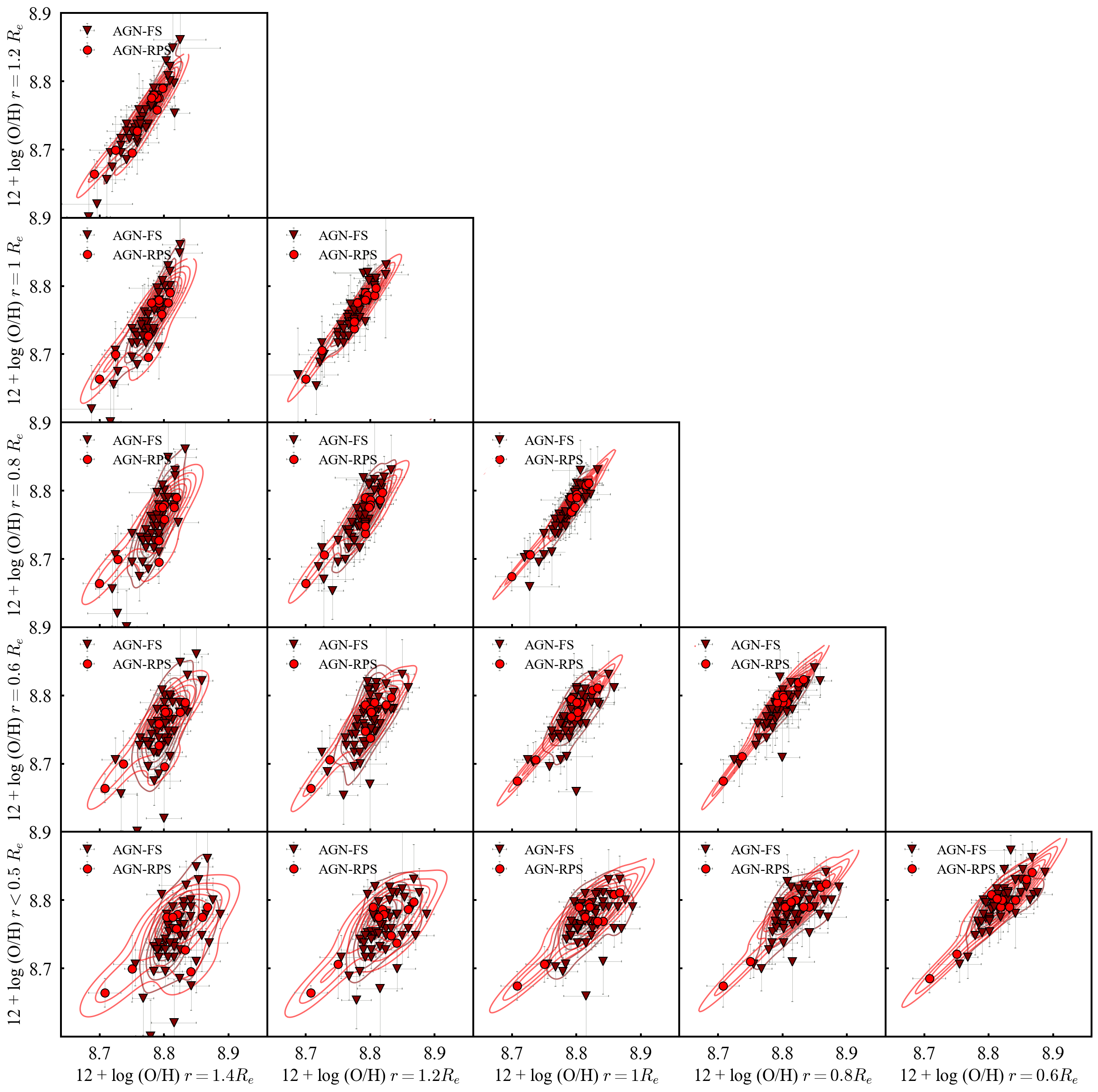}}%
    \caption{ Diagram comparing each pair of metallicities obtained at different radii for AGN-RPS(red circles)  and AGN-FS (dark red triangles). The lines show the probability density distribution evaluated using KDE.
    The comparison is performed among abundances computed at $r < 0.5 R_e$, $r = 0.6 R_e$, $r = 0.8 R_e$, $r = 1.2 R_e$ and $r = 1.4 R_e$.}
   \label{fig:results-Ia}
\end{figure*}


\subsection{
How the AGN $-$  RPS interplay affects the metallicity}

\begin{figure*}
    \centering
    \includegraphics[width=\linewidth]{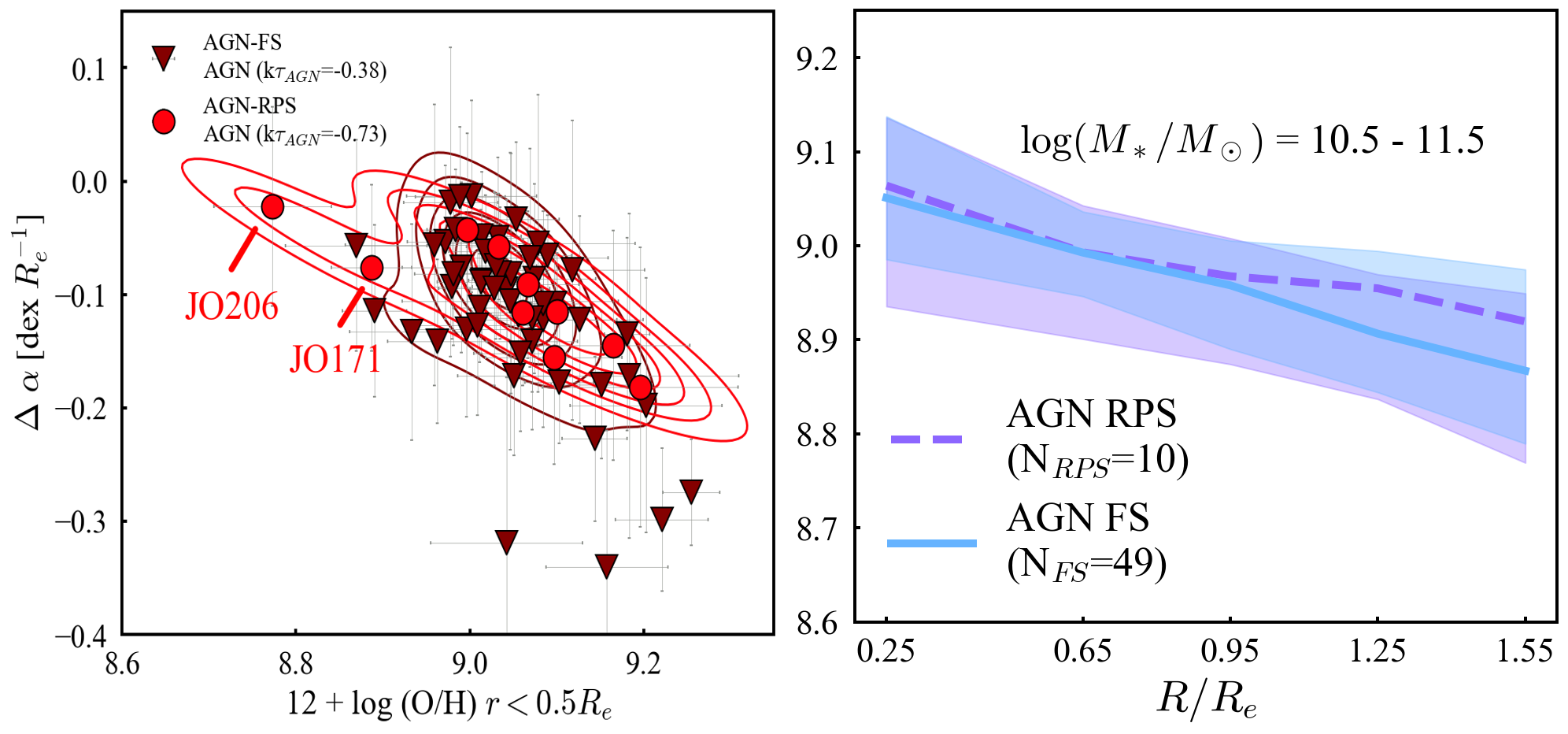}
    \caption{\emph{(Left Panel)} Gradient’s slope ($\Delta \alpha$) as a function of the nuclear metallicity. 
    The legend shows the Kendall $\tau$ test’s coefficients 
    revealing that the slopes seem to correlate with the nuclear metallicity only in the AGN-RPS,(k$\tau$AGN = -0.73). 
    \emph{(Right Panel)} Median oxygen abundance of the AGN-RPS galaxies (purple dotted line) and AGN-FS galaxies (blue continuous line), as a function of the galactic radius, spanning a mass range of $10.5 \leq \log M_*/M_\odot \leq 11.5$. The shaded areas cover the 25th/75th percentiles of the abundance's distribution inside each annulus.}
    \label{fig:AGN-FSvsRPS}
\end{figure*}



The complete atlas of the spatially-resolved gas-phase metallicity abundances of $477$ galaxies in our samples is published online\footnote{The online database is available at the link: XX}. 
As an example,  we show in Figure \ref{fig:gasp+manga} (in the first and second rows) the metallicity maps (on the left) and gradients (on the right) of the RP-stripped galaxy JO49 hosting an AGN ($\log M_*$= 10.68, $z=0.0451$) and the SF RP stripped galaxy JO93  ($\log M_*$ = 10.54, $z=0.037$), which does not show signs of AGN activity in its center. We pick these two galaxies since they have similar stellar mass. Both galaxies show steep and negative gradients. 
JO93 shows a plateau reaching the value 12 + log (O/H) $\sim$ 9.05 inside the region with radius $r \sim 0.5 \ R_e$. 
JO49, instead, shows a plume of points in the same region, which is in this case dominated by the AGN emission according to the BPT. As a consequence, the measured 12 + log (O/H) is 9.2.
We retrieve qualitatively similar gradients in the field, as shown in third and fourth rows of Fig.\ref{fig:gasp+manga} for the AGN galaxy `8311-6104’ ($\log M_*$ = 10.67, $z=0.027$) and the SF galaxy `8329-12701' ($\log M_*$ = 10.99, $z = 0.035$). 
Again, the gradient of the SF galaxy flattens at $r < 0.5 \ R_e$ around the value 12 + log (O/H) $\sim$ 9.0, while the abundances of the AGN galaxy increases steeply up to 12 + log (O/H) $\sim$ 9.2 in the same region, which is dominated by the AGN activity.
We note that in the SF galaxy 8329-12701, there is no detectable line emission in the outermost annuli, and because of this the gradient on the left panel shows the abundances up to $r \sim 1.5 R_e$. Similarly, the gradient of 8311-6104 extends up to the radius $\sim$ 2.2 $R/R_e$, because the last annulus ($r_{in} \sim$ 2.3 $R_e$ and $r_{out} \sim$ 2.5 $R_e$) contains less than 10 spaxels, which is the minimum threshold to compute the median 12 + log O/H. 

In the following, we consider the metal distribution of all the AGN in our sample, including both those hosted by ram-pressure stripped and undisturbed galaxies.
We compare the metallicity as a function of the galactic radius of the AGN-RPS and AGN-FS galaxies. Figure \ref{fig:results-Ia}  shows the distribution of the  RP-stripped AGN (red circles) and non-RP-stripped AGN (dark red triangles) galaxies in diagrams comparing the abundances inside concentric annuli at different galactocentric distances. We will use the term 'nuclear metallicity' to indicate the quantity 12 + log O/H$_{r<0.5 R_e}$, which is the median metallicity within the nuclear annulus with $r < 0.5 R_e$, and 'disk's metallicity' to indicate the abundance 12 + log O/H$_{r = 1 \ R_e}$, which is the median metallicity measured between $r= 0.9 \ R_e$ and $r= 1.1 \ R_e$. 

A 2D Kolmogorov Smirnov (2 KS) test run pairwise in the AGN-RPS and AGN-FS samples of each diagram is  not able to robustly establish whether the distributions are drawn from different parent samples.
We observe that the correlation in Figure \ref{fig:results-Ia} between the metallicity in the two most external annuli (e.g. 12 + log O/H$_{r = 1.2 R_e}$ versus 12 + log O/H$_{r = 1.4 R_e}$) is tight, as also the correlation between 12 + log O/H$_{r = 1 R_e}$ and 12 + log O/H$_{r = 0.8 R_e}$ in both the AGN-FS and AGN-RPS.
On the other side, the panels in the bottom row show that distributions tend to be more scattered when comparing the metallicity in the nuclear region ($R < 0.5 R_e$) with those in the outermost annuli.
This is indicative of a significant variation of the slopes.
Particularly, we show $\Delta \alpha$ as a function of the nuclear metallicity at $r < 0.5 \ R_e$ (12 + log O/H $_{r < 0.5 \ R_e}$) in Figure \ref{fig:AGN-FSvsRPS} (left panel). We note, first of all, that the slopes are all negative, ranging between [-0.3,0.0] dexR$_e^{-1}$.
According to a Kendall $\tau$ test (similar to a Spearman test, but better suited in case of small sample sizes), there is a strong correlation between $\Delta \alpha$ and 12 + log O/H $_{r < 0.5 \ R_e}$ among the AGN-RPS galaxies,  with a coefficient k$\tau_{AGN} = -0.73 $ (p-value = 0.8 $\times 10^{-3}$), while the AGN-FS does not show a significant correlation with the disk's metallicity within $r < 0.5 \ R_e$ (sp$_{AGN}$ = -0.37, p-value = 9 $\times 10^{-5}$). 

Two galaxies from the AGN-RPS sample, JO206 and JO171, stand out from the main distribution showing significantly flatter slopes at fixed nuclear metallicity, which also displays lower metallicities at any given radius as visible from Figure \ref{fig:results-Ia}. These two galaxies were also outliers in P23, being the only two AGNs lying below the SF mass-metallicity relation (SF MZR) (see Figure 7 of the paper). 
We describe some of the main properties of these outliers in Appendix \ref{sec:appendix-a}, even though a full comprehension of the reasons behind their peculiar metal distributions is beyond the scope of this work. 
Regardless of JO206 and JO171, the correlation in the AGN-RPS sample stands with a coefficient of k$\tau_{AGN} = -0.71$ (p-value = 0.01).
Interestingly, we also observe a significant scatter in 
four field AGN showing steeper gradients at a fixed value of 12 + log O/H$_{r < 0.5 R_e}$ than the rest of the AGN sample. In Appendix \ref{sec:appendix-a}, we show that these galaxies also have steeper slopes than those with the same nuclear luminosity and stellar masses.
According to the 2KS test, there is no significant difference in the distribution of the gradients in AGN-RPS and AGN-FS galaxies. 
Overall, the RPS seems to play a negligible effect in the metallicity gradients in our AGN sample. 
However, we stress that larger samples of RP-stripped galaxies hosting AGN activity are highly required to study the AGN-RPS connection further.
As an ultimate check, we compute the median abundances at a given $R/R_e$ among the 10 AGN-RPS and 52 AGN-FS galaxies (shown in the right panel of Figure \ref{fig:AGN-FSvsRPS}) and we observe that the maximum difference is
$\sim$ 0.05 dex,  well below the errors of $\sim$ 0.2 dex.
In conclusion, RPS does not play a role in regulating the metal content in AGN hosts as the metallicities at different radii do not differ from those in the field in the same stellar mass range.



\begin{figure*}[ht]
    \centering
    \makebox[\textwidth]{
    \includegraphics[scale=0.42]{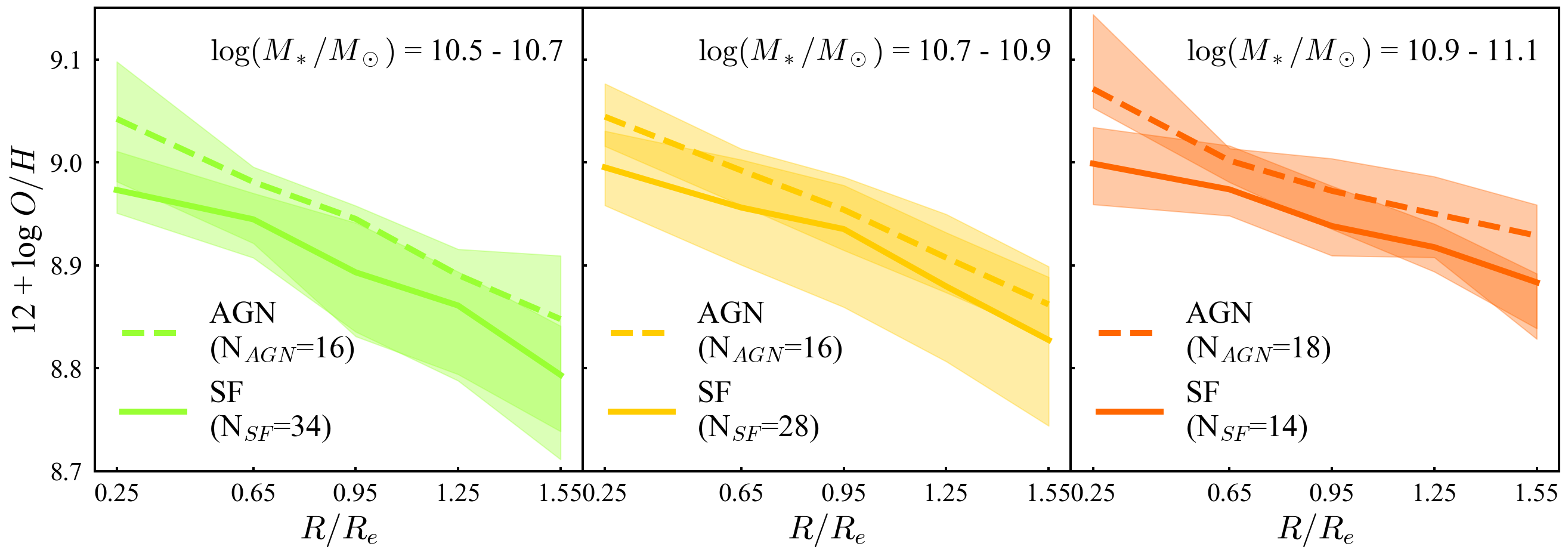}}%
    \caption{Median metallicity among AGN host galaxies (dashed line) and SF galaxies (continuous line) as a function of the galactic radius. The numbers (N$_{AGN}$, N$_{SF}$) of the galaxies in the two samples are reported in the legend for each mass bin. Shaded areas cover the upper and lower errors on the metallicity gradients, computed as the 25th/75th percentiles of the 12 + log O/H distribution inside each stellar mass bin. 
    As expected from the well-known MZR, the nuclear metallicity of both SF and AGN host galaxies constantly increases together with the host galaxy's stellar mass, saturating at 12 + log O/H $\sim$ 9.0 for the SF galaxies and rising to $\sim$ 9.1 in the AGN hosts. In addition, the metallicity gradients become flatter with higher stellar mass in the SF galaxies.
    }
    \label{fig:median_gradients}
\end{figure*}

\subsection{
Oxygen abundances of AGN hosts and SF galaxies} \label{sec:results-II}

In this section, we want to compare the abundances in SF and AGN host galaxies.
For completeness, we checked that the metallicity gradients of the stripped and non-stripped SF galaxies are also consistent, as for the AGN hosts (see Section \ref{sec:result}), within the errorbars, in a very wide stellar mass range ($9.0 \leq \log M_*/M_\odot \leq 11$).



For the reasons above, from now on we will not make any further distinction between field and RP-stripped galaxies, but we will focus on the comparison between AGN host galaxies versus pure SF galaxies without AGN activity. 
Since AGN hosts have stellar masses $\log M_*/M_\odot \geq 10.5$ 
we select the 83/412 SF galaxies, from the SF-FS+SF-RPS samples, above the same threshold. 

\begin{figure*}[ht]
     \centering
    \makebox[0.8\textwidth]{
    \includegraphics[scale=0.6]{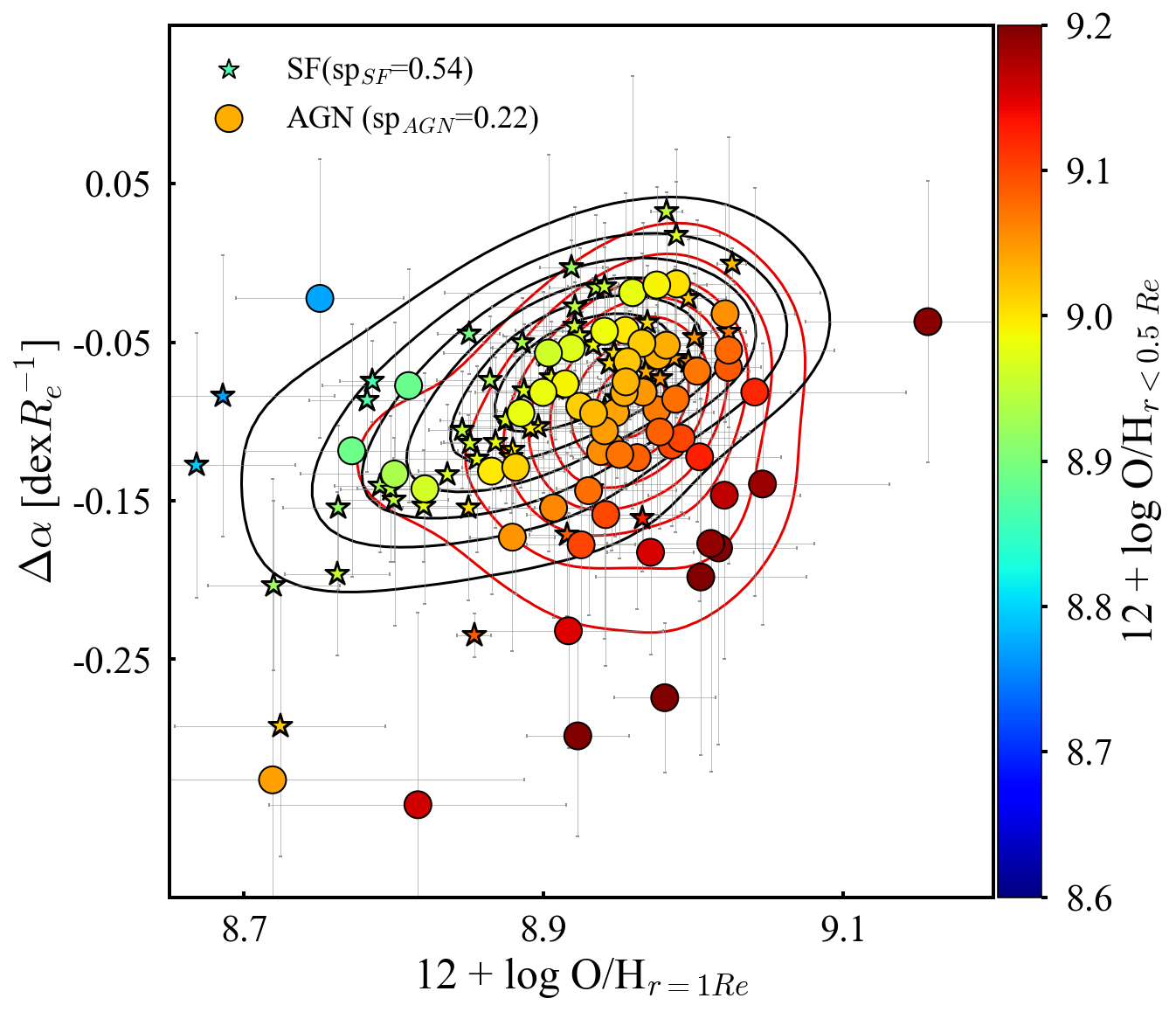}}%
    \caption{ Gradient's slope ($\Delta \alpha$) as a function of the disk metallicity 
    of SF galaxies (stars) and AGN host galaxies (circles) with $ \log M_*/M_\odot > 10.5$, color-coded according to the nuclear metallicity. 
    Black and red contours represent density curves computed by a KDE. In the legend, we also report the Spearman rank-order correlation coefficient that supports the finding that 
    the SF gradients strongly correlates with the disk's metallicity (sp$_{SF}$ = 0.54) but the same is not true for the AGN host galaxies. Indeed, AGN hosts have on average steeper slopes than SF due to their higher nuclear metallicities.}
    \label{fig:results-II}
\end{figure*}

Figure \ref{fig:median_gradients} shows the median oxygen abundance as a function of the galactocentric distance (expressed in units of effective radius, $R_e$) of AGN (dotted lines) and SF (continuous lines) galaxies, in three different mass bins. 


Accordingly to the plateau typically observed in the MZR of SF galaxies \citep[see e.g.][]{peluso+2022,mannucci+2010,tremonti+2004}, the nuclear metallicity of SF galaxies saturates at 12 + log O/H$_{r<0.5R_e} \sim$ 9.0, while the nuclear metallicity of the AGN host rises from 12 + log O/H$_{r<0.5R_e} \sim$ 9.05 (left and central panels) to 12 + log O/H$_{r<0.5R_e} \sim$ 9.1 (right panel), without showing signs of saturation. 
Also, AGN hosts show higher abundances than SF galaxies at any given radius. We  define the quantity:

\begin{center}
\Large
$R = \frac{{\rm 12 + \log O/H}_{r<0.5 R_e,{\rm AGN}} - {\rm 12 + \log O/H}_{r<0.5 R_e,{\rm SF}}}{{\rm 12 + \log O/H}_{r \sim 1.25 R_e,{\rm AGN}} - {\rm 12 + \log O/H}_{r \sim 1.25 R_e,{\rm SF}}}$
\end{center}

which is the ratio of the differences in metallicity between the AGN hosts and SF galaxies in the nuclear regions and in the disk at $r \sim 1.25 R_e$. Therefore, $R > 1$ means that the metallicity enhancement with respect to SF galaxies in the nuclear regions of AGN hosts is more accentuated than the enhancement observed in the disk. 
For both  $10.5 \leq \log (M_*/M_\odot) \leq 10.7$ and $\log (M_*/M_\odot) \geq$ 10.9, $R$ is $\sim 2.24$; at intermediate masses it is 
1.77. 
In other words, the AGN show stronger metal enhancement in their nuclear regions with respect to the one in the disk, regardless of the galaxy's stellar mass.

In Figure \ref{fig:results-II}, we plot the gradient's slopes ($\Delta \alpha$) versus the disk's metallicity, of both AGN hosts (circles) and  SF galaxies (stars), regardless of  environment and stellar mass. The points are color-coded according to the nuclear metallicity.
The black and red lines show the KDE of the SF and AGN host galaxies, respectively.  
We notice that the slopes are, again, predominantly negative in both the SF and AGN samples.

It now appears clear (as already hinted in Figure \ref{fig:median_gradients}) that the metal enhancement of AGN hosts galaxies is stronger than in SF galaxies in the nuclear regions.
At a fixed disk's metallicity, AGN hosts show steeper slopes than SF galaxies, with the  steepness increase  linked to their higher nuclear ($r<0.5 R_e$) metallicities.
Finally, using a Spearman test we found a significant correlation
between the slope $\Delta \alpha$ and the disk's metallicity in SF galaxies (sp$_{\rm SF}$ = 0.54, with a p-value of 2.3 $\times 10^{-10}$).
In contrast, AGN hosts do not show a correlation between the slope and 12 + log O/H$_{r = 1 \ R_e}$, by having a sp$_{\rm AGN}$ of 0.22 (p-value = 1.8 $\times 10^{-2}$).

\section{Discussion} \label{sec:discussion}

Thanks to the spatially-resolved oxygen abundance maps presented in this work, we were able to observe that AGN host galaxies show higher abundances than SF galaxies at each radius, but that the metal enhancement of the gas in the AGN's nuclear regions is accentuated with respect to what observed in the {\sc Hii} regions of the galactic disk.
In the following, we check for the presence or lack of correlation between the gradients' slopes and the host's galaxy properties. Also, we investigate a possible dependence of the results on the method adopted to compute the nuclear abundances, by comparing our results with those presented in \cite{donascimento+2022}, using a sample of galaxies in common with this work. We also test the robustness of our metallicity estimates by comparing the oxygen abundances obtained from the SEL calibrators in Section \ref{sec:calibrators} with those retrieved with calibrators and models from the literature.

\begin{figure*}[ht]
     \centering
    \makebox[\textwidth]{
    \includegraphics[scale=0.5]{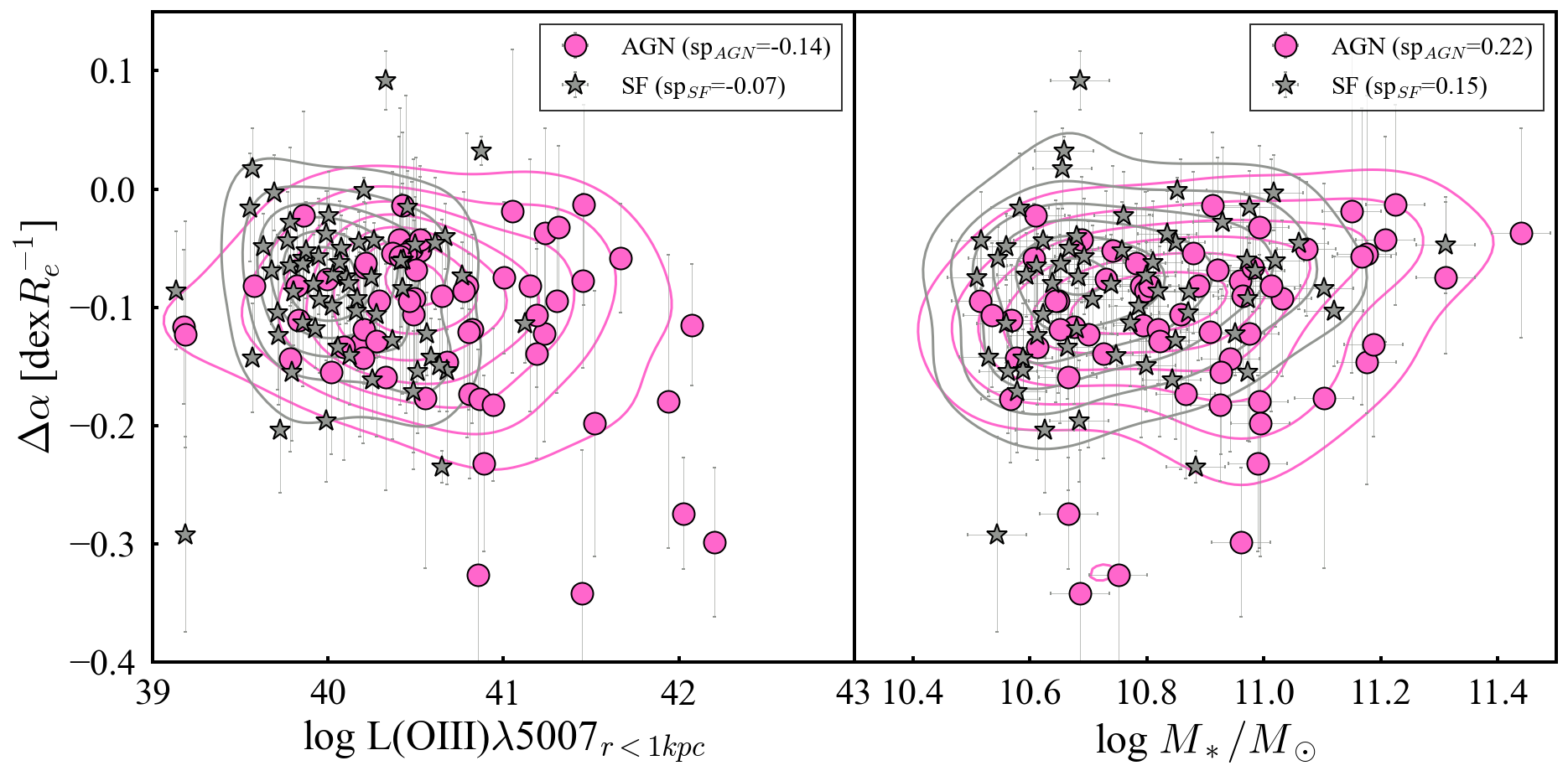}}%
    \caption{Slopes of the metallicity gradients ($\Delta \alpha$) of the AGN (pink points) and SF (gray stars) galaxies as a function of the luminosity of the \oiii~$\lambda$5007 line within an aperture with $r \sim$ 1 kpc ($\log L$ \oiii $\lambda$5007$_{r< 1 kpc}$) and the host galaxy stellar mass. Pink and gray contours represent density curves computed by a KDE. No clear correlation is observed in both the $\Delta \alpha$ - L \oiii $\lambda$5007 diagram  and $\Delta \alpha$ - $\log M_*/M_\odot$ diagrams in the SF and AGN samples, as supported by the Spearman rank-order correlation coefficient reported in the legend.
    } 
    \label{fig:AGNvsSF_discuss}
\end{figure*}

\subsection{SLOPES VERSUS GALAXY'S PROPERTIES}
In agreement with an inside-out growth scenario of the spiral discs through the accretion of cold gas streams \citep{matteucci+1989, molla+1996, boissier+1999, belfiore+2019a}, in the previous sections we have shown that in both AGN and SF galaxies the slopes of the metallicity gradients are mainly negative and range between [0,-0.3] dexR$_e^{-1}$.
To investigate the possible causes of the metal enhancement in the AGN-dominated nuclear regions with respect to those of SF galaxies (see Figure \ref{fig:results-II}), we show in Figure \ref{fig:AGNvsSF_discuss} $\Delta \alpha$ of the AGN (pink circles) and SF (gray stars) galaxies, regardless of their environment, as a function of the \oiii~$\lambda$5007 luminosity  ($\log$L\oiii$\lambda$5007$_{r < 1 kpc}$) and of the host galaxy's stellar mass, respectively. We consider the L \oiii~ in the NLR (which in this case we approximate to be the region within $r < 1$kpc) as a proxy of the AGN's bolometric luminosity \citep[see e.g.][]{berton+2015}.
 The Spearman coefficients reported in the legend of Figure 8 suggest that no correlation exists
between these two quantities and the gradient's slopes, in both our SF and AGN samples. 
The main difference between the two samples consists in the fact that  SF galaxies  reach at most L\oiii~$\sim 10^{41}$ (erg/s), while AGN host galaxies exceed this threshold. 
The right panel of Fig. 8 shows that not even a correlation between stellar masses and slopes is in place.
We checked that the nuclear luminosity does not scale linearly with the host galaxy's stellar mass, thus the SF and AGN relations in the $\Delta \alpha$ versus $\log M_*/M_\odot$ are not in contrast with those in the $\Delta \alpha$ versus $\log$ L \oiii$\lambda$5007$_{r < 1 kpc}$ diagram. 
We note that a residual dependence on the spatial resolution of the observation can affect these trends, since in the case of moderately luminous AGN (such as Seyfert galaxies) the NLR extends from $\sim$ 10 pc to $\sim$ 1 kpc \citep{ramos-almeida+2017} and, therefore, it is not always resolved in our samples.
To confirm the validity of these results, in Appendix \ref{sec:appendix-b} we also re-obtain the very well-studied relation between stellar mass and slopes of the SF gradients, confirming previous studies  \citep[e.g.,][]{mingozzi+2020,belfiore+2017b,franchetto+2020,khoram+2024b}. 
These trends are independent of the method adopted to measure the metallicity. 
In fact, \cite{mingozzi+2020} and  \cite{franchetto+2020} both use MAPPINGS IV \citep{dopita+2013} models but coupled with the code IZI \citep{blanc+2015} in the former and with a modified version of PYQZ \citep{dopita+2013} in the latter, while \citep{belfiore+2017b} exploit the calibrations from \cite{pettini-paguel+2004} and \cite{maiolino+2008}, as opposed to this work in which we rely on SEL calibrators (Section \ref{sec:calibrators}) based on {\sc Cloudy} v17 models coupled with the {\sc Nebulabayes} code.




\subsection{LINEAR FITTING OF THE STAR-FORMING DISK}

\begin{figure*}
\makebox[\textwidth]{
\includegraphics[scale=0.45]{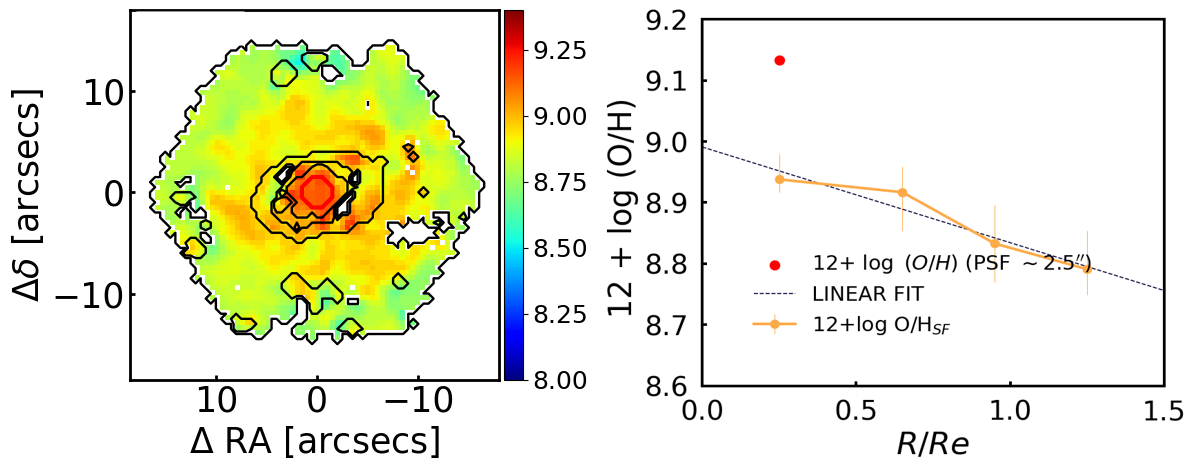}}%
\caption{(\emph{Left panel}) Metallicity map of the AGN-host galaxy `8985-12703' in MaNGA, which is a galaxy in common with the sample studied in \cite{donascimento+2022}. The red circle is the MaNGA PSF, e.g. an aperture with diameter $d \sim 2.5 ^{\prime \prime}$.
(\emph{Right panel}) Metallicity gradient for the same galaxy. The yellow line traces the median metallicity gradient if only SF spaxels are considered, while the red dot is the median 12 + log O/H inside the MaNGA PSF among all the spaxels (AGN/Composite/SF). Thus, the red dot traces the metallicity in the AGN-dominated region, while the orange gradient shows the metal content of the {\sc Hii} regions in the galaxy's disk.
The black dotted line is the linear fit of the observed {\sc Hii} gradient, obtained with an non-linear least square method. This figure is adapted in order to resemble Figure 3 in \cite{donascimento+2022}.
}
\label{fig:linear_fit}
\end{figure*}

\begin{figure}
\centering
\includegraphics[scale=0.4]{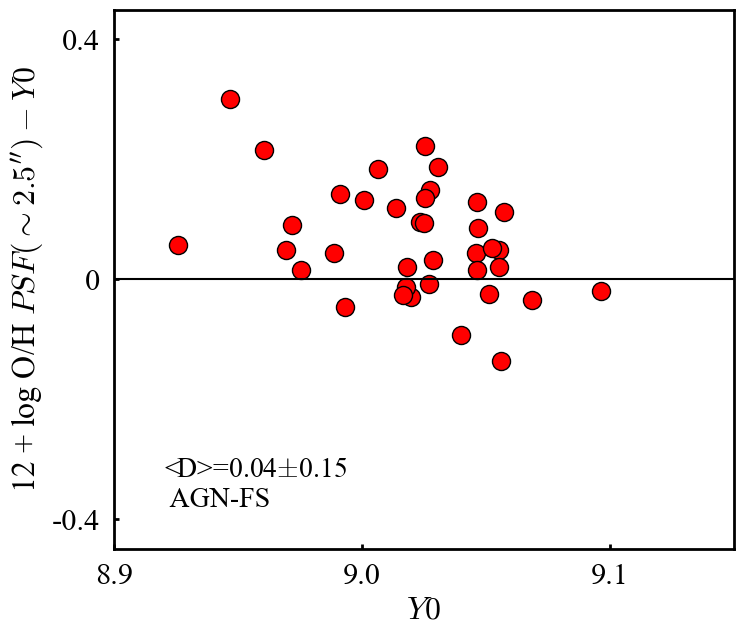}
\caption{Difference between the median oxygen abundance inside the MaNGA PSF (12 + log O/H PSF $\sim 2.5 ^{\prime \prime}$, e.g. red dot in Figure \ref{fig:linear_fit}) and the oxygen abundance extrapolated at $R = 0$ (e.g. Y0) from the linear fit, as a function of Y0. 
The measured metallicity of AGN hosts is systematically higher than the one extrapolated by the linear fit of the {\sc Hii} regions in the galaxy's center. 
}
\label{fig:donascimento}
\end{figure}

To ensure that the results presented in Section \ref{sec:results-II} are independent of the method employed to compute the nuclear metallicities in galaxies, we also repeat the analysis by applying the approach of \cite{donascimento+2022} (DN22, hereafter).
To briefly recap, we not only observe negative gradients in AGNs, so that the nuclear ($R< 0.5 R_e$) regions of AGNs are more metal enriched than the {\sc Hii} regions in the galactic disk ($R > 0.5 R_e$), but also that the enrichment is more enhanced relative to the nuclear regions of a control sample of galaxies without AGN (
see Figure \ref{fig:median_gradients}). 
This is in contrast with the findings of DN22, who 
measure lower nuclear metallicities than those extrapolated from a linear fit of the observed SF abundances in the galactic disk, in a sample of AGN hosts drawn from MaNGA.

DN22 perform a detailed study where they linearly fit the metallicities inside the SF spaxels of a sample of 61 AGN hosts and 112 SF galaxies selected from the DR15 of the MaNGA survey. The metallicity in the SF regions was computed using the \cite{pettini-paguel+2004} and \cite{perez-montero+2019} calibrators. The extrapolated value of the gradient to the galactic center ($R=0$) was then compared to the median value of 12 + log O/H inside an aperture with a diameter of 2.5'' (corresponding to a physical region extending from $\sim$1 to $\sim$6 kpc according to the galaxy's redshift), computed by using the \cite{storchi-bergmann+1998} and \cite{carvalho+2020} calibrators.
In DN22, the Composite spaxels were discarded.

To apply this kind of analysis to our AGN-FS sample drawn from MaNGA,
we fit as in DN22 the SF regions in our sample of 52 AGN hosts with the relation:

\begin{center}
Y = $Y_0$ + grad Y $\times$ R
\end{center}

 where Y is a given oxygen abundance -- in units of 12 + log O/H -- R is the galactocentric distance (in units of arcsec), $Y_0$ is the extrapolated value of the gradient to the galactic center ($R=0$), and grad Y is the slope of the distribution (in units of dex/arcsec).
 As an example, Figure \ref{fig:linear_fit} shows the linear fit of the gradient in the {\sc Hii} regions (yellow/orange line) of the MaNGA galaxy '8985-12703'. The red dot marks the median oxygen abundance inside an area equal to the MaNGA PSF (e.g. 12 + log O/H $_{PSF\sim 2.5 ^{\prime \prime}}$), which is the smallest resolved region.
 The linear fit is computed with a non-linear least square method for 42 galaxies out of 52 AGN hosts part of the AGN-FS, while for 10 galaxies the fit was not able to converge to a solution.  Figure \ref{fig:donascimento} displays the difference between 12 + log O/H $_{PSF\sim 2.5 ^{\prime \prime}}$ and Y0 extrapolated from the linear fit of the 42 AGN-FS galaxies. The median difference is compatible with zero, as indeed $<D>  = 0.04 \pm 0.15$. As opposed to this, the value found by DN22 is $-0.16$ (or $-0.30$ dex, depending on the calibration assumed for the AGN regions).
 In this work, we find that the metallicity in the AGN-dominated region is slightly higher than expected from pure star formation in galaxies with stellar mass $\log M_*/M_\odot \geq 10.5$. 
 The difference becomes significant, $<D> = 0.10 \pm 0.09$, and only four galaxies have $<D> \ \leq 0$ if we consider only the more massive AGN hosts ($\log M_*/M_\odot \geq 10.9$), which show signs of metal pollution induced by the AGN activity, as observed in the right panel of Figure \ref{fig:median_gradients}.


 
DN22 find the opposite result, which is that the measured values with the \cite{storchi-bergmann+1998} and \cite{carvalho+2020} calibrators are systematically lower than the metallicity at $R=0$ extracted from the linear fit of the {\sc Hii} regions.
In other words, in DN22 the measured nuclear AGN-dominated metallicity is lower than the one expected from a scenario in which star formation is the dominant ionization mechanism at the galaxy's center.


The discrepancy between the findings in this work and in DN22, thus, may depend on the fact that DN22 
used calibrators obtained by different and independent works, while we use calibrators 
based on 
consistent assumptions among the AGN, Composite, and SF models and using the same code (see Sec. \ref{sec:result}). 


In addition, it is also important to keep in mind that the classification of AGN and SF regions is based on different criteria: while we used \nii-BPT diagram, DN22 relied on the WHAN diagrams proposed by  \cite{cid-fernandes+2011}, which are based on the H$\alpha$ equivalent width and log \nii/H$\alpha$.


In conclusion, we find that the nuclear regions of a galaxy are more metal-enriched when AGN activity is present, independently from the method adopted to obtain nuclear metallicity.

\subsection{COMPARISON WITH AGN- AND SF- CALIBRATORS FROM THE LITERATURE}

\begin{figure}
    \centering
    \includegraphics[scale=0.65]{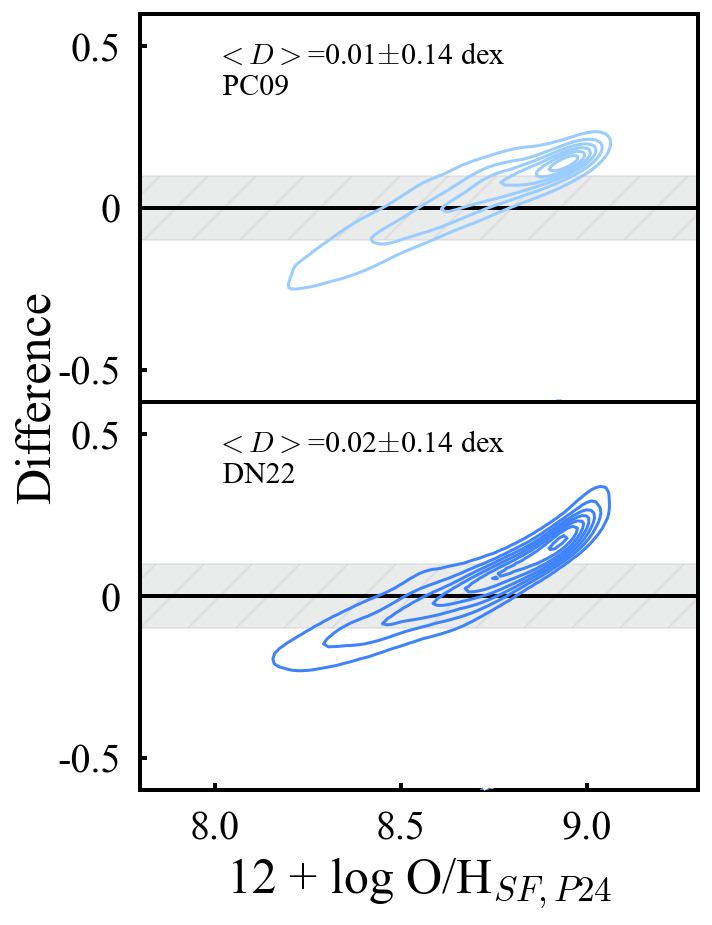}
    \caption{Difference between 12 + log (O/H) inside the SF spaxels of the SF-RPS and SF-FS galaxies computed with the P24 calibrator in Equation \ref{eq:sf_calib} (12 + log O/H$_{SF,P24}$) and the P09 (top panel) and DN22 (bottom panel) calibrators, as a function of 12 + log O/H $_{SF,P24}$. The median difference, $<D>$, and the standard deviation ($\pm \sigma$) of the distribution are reported in the legend.}
    \label{fig:comp_SFcalibrators}
\end{figure}

\begin{figure}
    \centering
    \includegraphics[scale=0.65]{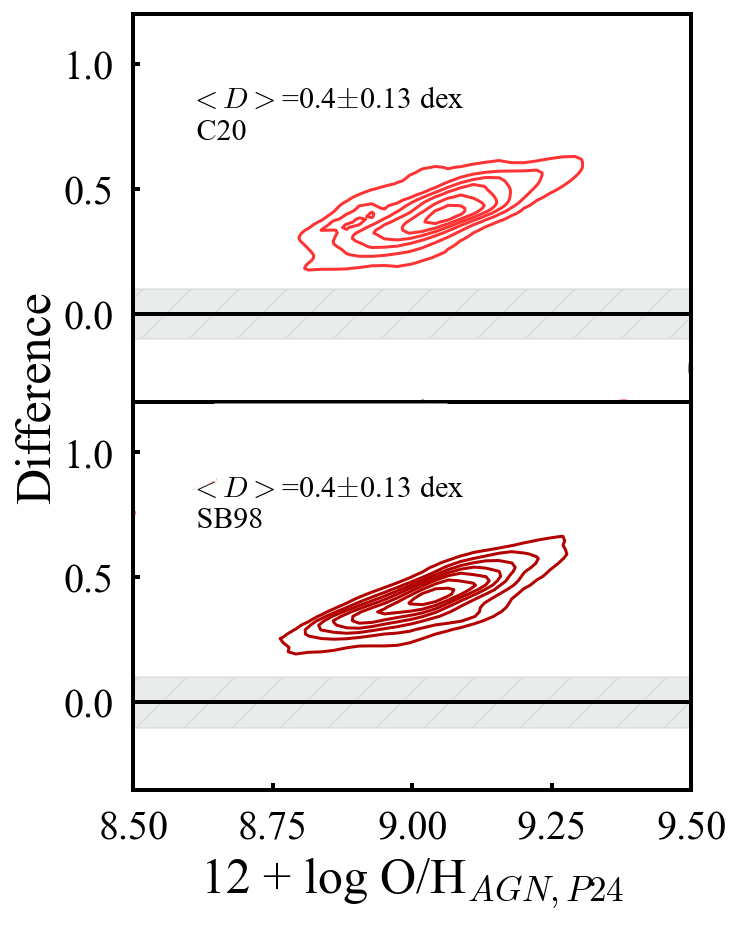}
    \caption{Difference between 12 + log (O/H) inside the AGN spaxels of the AGN-RPS and AGN-FS galaxies computed with the P24 calibrator in Equation \ref{eq:agn_calib} (12 + log O/H$_{AGN,P24}$) and the C20 (top panel) and SB98 (bottom panel) calibrators, as a function of 12 + log O/H$_{AGN,P24}$. The median difference, $<D>$, and the standard deviation ($\pm \sigma$) of the distribution are reported in the legend.}
    \label{fig:comp_AGNcalibrators}
\end{figure}

\begin{figure}
    \centering
    \includegraphics[scale=0.6]{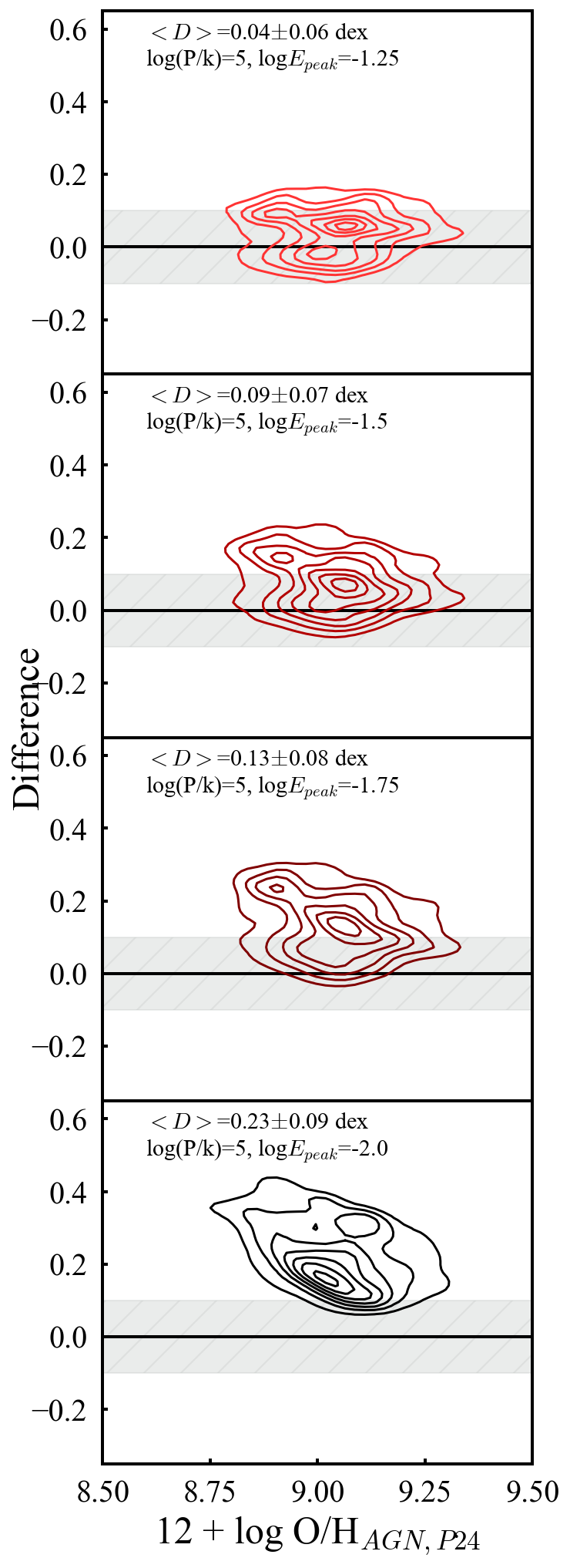}
    \caption{Difference between 12 + log O/H$_{AGN, P24}$ and the values obtained giving in input to {\sc NebulaBayes} the \oiii/\sii and \nii/\sii~ ratios from {\sc MappingsV} models \citep{thomas+2018a} with a peak energy in the AGN accretion disk of $\log E_{peak}$ = -1.25, -1.5, -1.75 and -2.0. The gas pressure is fixed to $\log P/k$ = 5, after checking that the results do not change as a function of this parameter. The grey shaded area covers the range $\pm$ 0.1 dex, which is the uncertainty associated to the parameters estimates with NebulaBayes. The median difference, $<D>$, and the standard deviation ($\pm \sigma$) of the distribution are reported in the legend.}
    \label{fig:comparison_mappings}
\end{figure}

To further strengthen our analysis and results, in this section, we re-compute again the metallicity for all our galaxies using SEL calibrators from the literature and compare the outputs to our estimates obtained with the {\sc SF} and {\sc AGN} calibrators in Equations \ref{eq:sf_calib} and \ref{eq:agn_calib}.

  To compute the abundances in the SF-classified spaxels, we use the calibrator given in  Equation (6) of DN22  and the calibrator taken from \citet{perez-contini+2009} (PC09, hereafter) and based on direct $T_e$ method's estimates. 
  
 To compute the abundances in the AGN-classified spaxels, we consider the calibrators from \cite{storchi-bergmann+1998} (SB98, hereafter) and \cite{carvalho+2020} (C20, hereafter), as indeed these equations involve strong emission lines occurring within the MUSE spectral coverage. 
 Table \ref{tab:calibs} summarizes the main assumptions made for the calibrators and models retrieved from the literature, particularly: log(N/O) calibration with 12 + log(O/H) in the regime of secondary N production, the shape of the AGN ionizing continuum, the gas density and solar oxygen abundance. 
For comparison, we also recap the assumptions made to derive the AGN calibrator presented in this work.

The DN22 calibrator is:


\begin{equation}
12 + \log({\rm O/H}) = 9.07 + 0.79 \times N2
\end{equation}

where N2=$\log$\nii$\lambda$6584/H$\alpha$ ratio, and the P09 calibrator is:

\begin{equation}
12 + \log({\rm O/H}) = 8.74 - 0.31 \times O3N2
\end{equation}

where O3N2 \citep{alloin1979} is defined as:

\begin{equation}
  O3N2 = \log \frac{\oiii\lambda 5007}{H\beta}
 \small{\times} \frac{(H\alpha)}{\nii\lambda6584}
\end{equation}

These relations are valid for 12+ $\log {\rm O/H}$ $\geq$ 8.0.
Figure \ref{fig:comp_SFcalibrators} shows the KDE of the difference between abundances obtained with Equation \ref{eq:sf_calib} (e.g., 12 + log O/H$_{P24,SF}$) and with the P09/DN22 calibrator (top/bottom panel) as a function of 12 + log O/H$_{P24,SF}$. We color in grey the area spanning a difference between $\pm$ 0.1 dex, which is the typical uncertainty assumed for the metallicity estimates through SEL calibrators.
We find that the standard deviation of the distribution is $\approx$ 0.14 dex for both the PC09 and DN22 calibrators, meaning that the difference between our estimates and those from PC09/DN22 is $< 0.14$ dex for the majority (50\%) of the points.   

As far as the  AGN calibrator is concerned, we apply the following SB98 equation: 

\begin{center}
    $12 + \log {\rm O/H}_{SB98} =  8.34 + 0.212x - 0.012x^2 - 0.002y +0.007xy$ \\ $- 0.0002x^2y + 6.52 \times 10^{-4} y^2 + 2.27 \times 10^{-4} xy^2$ \\ $+ 8.87 \times 10^{-5} x^2y^2$
\end{center}
\begin{equation}
\end{equation}

where x $\equiv$ (\nii$\lambda$6548 + \nii$\lambda$6584)/H$\alpha$ and y $\equiv$  (\oiii$\lambda$4959 + \oiii$\lambda$5007)/H$\beta$. 
 SB98 uses CLOUDY  to predict fluxes in case of a segmented power-law radiation field described in \cite{Mathews1987}. 
The model oxygen abundance ranges between 8.4 $\leq$ 12 + log (O/H) $\leq$ 9.4, and the ionization parameter -4.0 $\leq$ log (U) $\leq$ -2.0. 
They account for dust depletion and assume secondary-origin nitrogen following \cite{storchi-bergmann+1994}. 
The gas density is fixed to $n_{H}$ = 300 cm$^{-3}$. 


We also adopt the SEL calibrator by C20:  

\begin{equation}
    12 + \log {\rm O/H}_{C20} = 8.69 +  4.01^{{\rm N2}} -0.07
\end{equation}

The validity range is -0.7 $\leq$ N2 $\leq$ 0.6, which translates into a validity range between 8.16 $\leq$ 12 + log O/H $\leq$ 9.0.
This relation is obtained by interpolating CLOUDY v17.0 \citep{ferland+2017} photoionization models to the observed $\log$ (\oiii$\lambda$5007/\oii$\lambda$3727) versus $\log$ (\nii$\lambda$6583/H$\alpha$) in a sample of 433 Seyfert from the SDSS DR7 \citep[Sloan Digital Sky Survey]{abazajian+2009} and selected by \cite{dors+2020a}. 


\begin{table*}
\begin{tabular}{c  p{5cm}  p{4cm} p{2.5cm} p{3.5cm}}
Authors & Nitrogen secondary-origin   &  Ionizing continuum & gas-density/\newline gas pressure & solar abundance \newline \\         
\hline 

\bigskip
\\
SB98 & $\log {\rm N/O}$ = 1.29 $\times$   
(12 + $\log {\rm O/H}$ - 11.84) \newline \citep{storchi-bergmann+1994} &  segmented power law with $\alpha$ = -1 and -1.5 \newline \citep[turn-on energy at 0.1 ev, turn-off energy at 100 keV,][]{Mathews1987} & $n_{H}$ = 300 cm$^{-3}$ & 12+log (O/H)$_\odot$=8.9 \newline \citep{aller1987} \newline \citep{grevesse+1984} \\ 
\\
\hline
\bigskip
\\
C20 & $\log {\rm N/O}$ = 0.96 $\times$ 
(12 + $\log {\rm O/H}$ - 9.29) \newline  \citep{dors+2017}  &  slope $\leq$ -1.4 between 2 keV and 2500 \AA~
& $n_{H}$ = 100, \newline 300, \newline 500 cm$^{-3}$ & 12+log(O/H)$_\odot$= 8.69 \newline \citep{asplund+2009} \\
\\
\hline
\bigskip
\\
T18 & $\log {\rm N/O}$ = -1.73 $\times$ (log O/H + 2.19)  \newline \citep{nicholls+2017} & NLR ionizing spectrum with the energy of the peak of the accretion disk emission $-2.0 \geq \log E_{peak} \geq -1.25$, \newline photon index of the inverse Compton scattered power-law tail,  $\Gamma$ = 2,  \newline and the proportion of the total flux that goes into the non-thermal tail, $pNT$ = 0.15 \citep{thomas+2016} & $\log(P/k)$ = 5 & 12+log(O/H)$_\odot$ = 8.76 \newline \citep{nicholls+2017} \\
\\
\hline
\\
This work & log (N/H) = -4.57 + log (Z/Z$_\odot$) for log (Z/Z$_\odot$) $\leq$ 0.63; \newline log (N/H) =  - 3.94 + 2 $\cdot$ log (Z/Z$_\odot$) otherwise \newline
\citep{dopita+2000} & power law continuum from {\sc table power law} command in \bm{{\sc Cloudy}}  with slope $\alpha =-2$ at $\nu \geq$ 3676 Ryd & $n_{H}$ = 100 cm$^{-3}$ & 12+log(O/H)$_\odot$ = 8.69  \newline \citep{asplund+2009} \\

\end{tabular}   
\caption{Summary of the main assumptions to derive the calibrators and models retrieved from the literature, used to make a comparison with Equation \ref{eq:agn_calib} describing the AGN calibrator presented in this work. Columns are: (i) relation for the nitrogen secondary origin (+reference paper); (ii) ionization continuum to model the emission from the AGN's accretion disk; (iii) gas density and/or gas pressure; (iv) reference value assumed for the solar abundance.}
\label{tab:calibs}
\end{table*}

Figure \ref{fig:comp_AGNcalibrators} shows the difference between the abundances in the AGN regions obtained by the different calibrators. 
In both cases, the difference between abundances from the P24 calibrators and from the C20/SB98 calibrators is $<$ 0.4 dex within $1 \sigma$ of the distribution. 
Also, the KDE distributions show that the difference is always higher than zero in both cases. In other words, the SB98 and C20 calibrators tend to underestimate the metallicities with respect to Equation \ref{eq:agn_calib}, and the difference linearly increases along with the metallicity. 

To further investigate this discrepancy, we also compare the abundances obtained with our models and with the {\sc MappingsV} \citep{sutherland-dopita+2017} photo-ionization models presented in \cite{thomas+2018a} (T18 models, hereafter). We apply the T18 models following the same procedure described in Section \ref{sec:method}, thus comparing predicted and observed \oiii/\sii and \nii/\sii. The T18 models are plane-parallel, one-dimensional, and dusty, with elemental depletions onto dust grains based on \cite{jenkins2009}. The assumed solar oxygen abundance is 12 + log O H = 8.76. The Seyfert ionizing spectrum is taken from \cite{thomas+2016} and parameterizes the energy of the ionizing accretion disk emission by its peak energy, $E_{peak}$, the photon index of the inverse Compton scattered power-law tail, $\Gamma$, and the proportion of the total flux that goes into the non-thermal tail, $pNT$.
The only parameter left to vary was $E_{peak}$. The other parameters are fixed to the fiducial values $\Gamma$ = 2.0 and $pNT$ = 0.15. 
 In Figure \ref{fig:comparison_mappings}, we show the difference between abundances computed with the T18's {\sc MappingsV} models having $\log E_{peak}$ (keV) = -1.25, -1.5, -1.75, -2.0 and 12 + log O/H$_{P24,AGN}$. We checked that the T18 models do not change significantly as a function of the gas pressure, which we set to log (P/k) = 5.  
 We observe generally good agreement up to $\log E_{peak}$ (keV) = -1.5, with the median difference, $<D>$, equal to 0.04$\pm$0.06 dex and 0.09$\pm$0.07 dex. The discrepancy starts to increase at $\log E_{peak}$ (keV) = -1.75 with $<D>$ = 0.13 $\pm$ 0.08 dex, and becomes significant for 
  $\log E_{peak}$ (keV) = -2.0, having $<D>$ = 0.23 $\pm$ 0.09 dex, which are higher than the 0.1 dex uncertainty associated to the {\sc NebulaBayes} parameters estimate (see Figure \ref{fig:errors}).  \cite{thomas+2019} adopt an Epeak = 45 eV, corresponding to log Epeak (keV)  $\sim$ - 1.35, as values of Epeak in the range 40–50 eV resulted to be more plausible \citep{thomas+2018b}.
In the case of a harder AGN radiation field, we observe that the metallicities decrease and get closer to the values obtained by SB98 and C20. However, as explained in detail in \cite{thomas+2018a}, the extreme value of $E_{peak}$ = 100 eV (e.g. $\log E_{peak}$ = -2.0) is improbably high considering the ionization potentials of species observed in typical NLRs. 



\section{Conclusions} \label{sec:summary}


In this work, we explore how the interplay between RPS and AGN activity affects oxygen abundance throughout the galaxy, as well as the influence of AGN activity on the metallicity in the galactic nuclear regions.

To measure the metallicity, we use SEL calibrators, referred to as P24 calibrators, obtained by linearly fitting {\sc Cloudy} photoionization models generated assuming ionization from stars, AGN, and a mix of both. With these models, we reproduce the line ratios \nii/\sii~ and \oiii/\sii~ to constrain the oxygen abundance and the ionization parameter by matching predictions and observations with the code {\sc NebulaBayes}. 
To validate the use of our new set of SEL calibrators, we compare the P24 calibrators with widely used calibrators retrieved from the literature.

First, we measure the spatially resolved gas-phase metallicity of 10 AGNs experiencing RPS from the GASP survey and 52 undisturbed field AGNs from the MaNGA DR15.
We find that:

\begin{itemize}
    \item[(i)] the metallicity distributions of stripped and non-stripped AGN are statistically indistinguishable within r $< 1.5 R_e$, with the only exception of two RPS galaxies, JO206 and JO171, which show lower (see Figure \ref{fig:AGN-FSvsRPS}) abundances at any given radius than the rest of the AGN-RPS's and AGN-FS's galaxies.
\end{itemize}

The above finding shows that the synergy between AGN and RPS does not play a major role in altering the gas-phase metallicity in the GASP galaxies.
This result aligns with the fact that the GASP sample consists of stripped galaxies likely experiencing the peak intensity of RPS. This is supported by their enhanced SFR with respect to main sequence galaxies \citep{vulcani+2018},
which is expected during the peak intensity phase of RP \citep{ricarte+2020}. In addition, they feature remarkably long gas tails that extend for tens of kpc along with star-forming knots \citep{gullieuszik+2023,giunchi+2023}. According to simulations by \cite{ricarte+2020}, the combined effect of RPS and AGN feedback on the metallicity distribution becomes significantly effective and induce a decline of the SFR, and consequently of the metal pollution, by a factor of 10 over a few Myrs after the RP peak intensity.
Therefore, to better characterize the link between the two phenomena,  
we would need larger samples of RPS galaxies captured after the RP peak phase. However, we speculate that these observations would be extremely challenging due to the faintness of the gas emission in these low-SFR systems.

By including a control sample of 83 SF galaxies, we further explore the effect of the AGN activity on the metallicity distributions.
We compare the metallicity gradients of these galaxies with those of AGN in both clusters and the field, allowing us to analyze the effects independently of the galaxies’ environments. All galaxies considered have stellar masses of $\log (M_*/M_\odot) \geq 10.5$ and are observed at redshift $z< 0.07$.
We find that:

\begin{itemize}
     \item[(ii)] AGN hosts show higher metallicities than SF galaxies at any radius. However, the AGN hosts present a metal enrichment in the nuclear regions ($r < 0.5 R_e$) higher by a factor ranging between 1.8 and 2.3 times the enhancement shown in the disk at $r \sim 1.25 R_e$ (see Figure \ref{fig:results-Ia}) depending on the galaxy's stellar mass. Particularly, this factor is 2.3, 1.77, and 2.24 in the mass bins $ 10.5 \leq \log M_*/M_\odot \leq 10.7$, $ 10.7 \leq \log M_*/M_\odot \leq 10.9$ and $ 10.9 \leq \log M_*/M_\odot \leq 11.1$.
     \item[(iii)] In both SF and AGN galaxies, the slopes of the radial profile do not correlate significantly with other galaxy's properties, such as the nuclear \oiii~ luminosity ($\log$L\oiii$\lambda$5007$_{r < 1 kpc}$) and host galaxy stellar mass ($\log M_*/M_\odot$).
\end{itemize}


The enhanced metallicity observed in AGN can be explained by several hypotheses. One explanation is that the gas is enriched in metals from the dissociation and disruption of dust that pollutes the ISM in the SMBH's vicinity \citep{maiolino-mannucci2019}. 
Another possible explanation for the difference in metallicity distributions between SF galaxies and AGN hosts is the reduced SFR associated with the presence of an AGN rather than by the AGN itself \citep{li+2025,schawinski+2014,ellison+2016,lacerda+2020}. This reduced SFR could lead to higher metallicity in AGN hosts, as the relationship between SFR and metallicity is known to be anti-correlated according to the Fundamental Metallicity Relation \citep[FMZR,][]{mannucci+2010}.
\cite{li+2025} proposed two possible mechanisms to explain the suppressed SFR in AGNs.
One is a "gentle" AGN negative feedback that prevents gas accretion \citep{kumari+2021}. The other possibility is that the growth of the bulge simultaneously triggers both AGN activation and the inhibition of star formation. \\
\\
Overall, the above considerations and results emphasize the need for multi-frequency information about gas and dust across various ionization stages, with resolutions $<<$ 1 kpc. 
In fact, we want to stress once again that there may still be a residual and intrinsic dependence of our results on the spatial resolution ($\sim$ 1 kpc) of these observations.
Sub-kpc spatial resolutions are crucial for resolving the NLR and understanding its metal content, as well as for distinguishing between star formation occurring in the vicinity of AGN and within the galactic disk. An in-depth understanding of these properties is essential for linking observed oxygen abundances to the physical mechanisms underlying their formation.

\begin{acknowledgements}
Based on observations collected at the European Organization for Astronomical Research in the Southern Hemisphere under ESO program 196.B-0578. This project has been co-funded by the European Research Council (ERC) under the European Union's Horizon 2020 research and innovation program (grant agreement No. 833824). 
GP and ID acknowledge funding from the European Union – NextGenerationEU, RRF M4C2 1.1, Project 2022JZJBHM: "AGN-sCAN: zooming-in on the AGN-galaxy connection since the cosmic noon" - CUP C53D23001120006.
JF acknowledges financial support from the UNAM- DGAPA-PAPIIT IN110723 grant, México.

This work made use of the {\sc KUBEVIZ} software which is publicly available at \url{https://github.com/matteofox/kubeviz/}. The development of the KUBEVIZ code was supported by the Deutsche Forschungsgemeinschaft via Project IDs: WI3871/1-1 and WI3871/1-2.
This work makes use of data from SDSS-IV. Funding for SDSS has been provided by the Alfred P. Sloan Foundation and Participating Institutions. Additional funding toward SDSS-IV has been provided by the U.S. Department of Energy Office of Science. SDSS-IV acknowledges support and resources from the Center for High-Performance Computing at the University of Utah. The SDSS website is www.sdss.org. This research made use of Marvin, a core Python package and web framework for MaNGA data, developed by Brian Cherinka, José Sánchez- Gallego, and Brett Andrews \citep{cherinka+2019}. SDSS-IV is managed by the Astrophysical Research Consortium for the Participating Institutions of the SDSS Collaboration including the Brazilian Participation Group, the Carnegie Institution for Science, Carnegie Mellon University, the Chilean Participation Group, the French Participation Group, Harvard-Smithsonian Center for Astrophysics, Instituto de Astrofísica de Canarias, The Johns Hopkins University, Kavli Institute for the Physics and Mathematics of the Universe (IPMU)/University of Tokyo, Lawrence Berkeley National Laboratory, Leibniz Institut für Astrophysik Potsdam (AIP), Max-Planck-Institufür Astronomie (MPIA Heidelberg), Max-Planck-Institut für Astrophysik (MPA Garching), Max-Planck-Institut für Extra- terrestrische Physik (MPE), National Astronomical Observa- tory of China, New Mexico State University, New York University, University of Notre Dame, Observatário Nacional/ MCTI, The Ohio State University, Pennsylvania State University, Shanghai Astronomical Observatory, United King- dom Participation Group, Universidad Nacional Autónoma de México, University of Arizona, University of Colorado Boulder, University of Oxford, University of Portsmouth, University of Utah, University of Virginia, University of Washington, University of Wisconsin, Vanderbilt University, and Yale University.
The MaNGA data used in this work are publicly available at \url{http://www.sdss.org/dr15/manga/manga-data/}.  
\end{acknowledgements}

\clearpage

\appendix 

\section{Properties of the outliers among the stripped and non-stripped AGNs} \label{sec:appendix-a}

In this Appendix we present some of the fundamental properties of a sample of 'peculiar' galaxies, considered as such due to their significant scatter from the main distributions in Figures \ref{fig:results-Ia} and \ref{fig:AGN-FSvsRPS}.
However,  an in-depth analysis of the link between these properties and the peculiar metallicity distributions characterizing these galaxies is beyond the scope of this work. 

First, we focus on JO206 ($z=0.051$, $\log M_*/M_\odot$ = 10.96) and JO171 ($z=0.052$, $\log M_*/M_\odot$ = 10.61), two GASP galaxies whose metallicity is significantly lower than that covered by the rest of the sample (see Fig. \ref{fig:results-Ia}). 

We show in Figure \ref{fig:pecgal0} their spatially-resolved metallicity map, metallicity gradient, and map of the \oiii luminosity ($L$ \oiii$\lambda$5007) and $\log U$. 
By considering all the spaxels within a galacticentric distance of $r \sim 1$ kpc  (e.g. delimitated by the orange circles in the bottom panels of Figure \ref{fig:pecgal0}), we find that the median values inside this aperture are $\log$L\oiii$\lambda$5007$_{r < 1 kpc}$ = 41.45 and $\log U_{r < 1 kpc}$ = -2.56 for the galaxy JO206, and  $\log$L\oiii$\lambda$5007$_{r < 1 kpc}$ = 39.86 and $\log U_{r < 1 kpc}$ = -2.74 for the galaxy JO171. 
In Figure \ref{fig:logu-vs-logo3}, we plot  $\log$L\oiii$\lambda$5007$_{r < 1 kpc}$ as a function of $\log U_{r < 1 kpc}$ on the left panel and $\log$L\oiii$\lambda$5007$_{r < 1 kpc}$ as a function of the host galaxy's stellar mass for all the AGN-FS (dark-red triangles) and AGN-RPS (light-red circles), highlighting the 'peculiar' galaxies.
We observe that JO206 and JO171 do not stand out from the rest of the AGN-RPS sample in terms of their nuclear luminosity or strength of ionization radiation.
However, as hinted by its peculiar morphology, JO171 is a rare example of a ring galaxy perturbed by RPS \citep[see][]{moretti+2018}. A joint analysis of the stellar populations and the gas/stellar kinematics performed by \cite{moretti+2018} showed that the origin of the ring was presumably due to a gas accretion event that happened in the distant past, which was then followed by a stripping event consequently to the infall into the massive cluster Abell 3667. 
Other than that, a recent deep (200 ks) \emph{Chandra} program targeting JO206 has revealed the presence of an unusually high X-ray luminous AGN \citep[see e.g.][]{fotopouolou+2024}, with L$_{X}$ = 7-9 $10^{49}$ erg/s in the 0.5-7.0 keV band (Ignesti, private communication).   
The currently accepted scenario for the production of X-ray photons around SMBH includes the Comptonization of disk photons in a population of hot thermal electrons \citep{haardt_maraschi_1993}. The torus is responsible for the obscuration of X-rays due to photoelectric absorption.
The $\emph{Chandra}$ spectrum of JO206 reveals a Compton-thin (e.g. $\log N_{H} ({\rm cm}^{-2}) \sim 22$) and obscured (e.g. $\gamma \sim$ 1.4) hot corona, where the absorbed part of the spectrum is in the 0.5 $-$ 5 keV range.


Next, we focus on $4$ MaNGA galaxies which have steeper slopes than the main distribution of the AGN-FS, going up to $\alpha \sim$ 0.3 dex (see Figure \ref{fig:AGN-FSvsRPS}).  These are the AGN hosts '8993-12705' ($z=0.029$, $\log M_*/M_\odot = 10.96$), '8311-6104' ($z=0.027$, $\log M_*/M_\odot = 10.67$), 8456-3701 ($z = 0.026$, $\log M_*/M_\odot =$ 10.75) and 9027-12704 ($z = 0.035$, $\log M_*/M_\odot = 10.69$) 
and we show their 
spatially-resolved metallicity maps, metallicity gradients, maps of $\log$  L\oiii$\lambda$5007 and $\log U$ in Figure \ref{fig:pecgal0}. , 
The metallicity map of 8456-3701 reveals a highly dis-homogenous metal distribution which, joined to the irregular galaxy morphology, may suggest a strongly disturbed gas distribution residual of a past interaction. 
It is clear from Figure \ref{fig:logu-vs-logo3} that 2/4 galaxies display the highest values of $\log U$ $_{r < 1 kpc}$ among all the AGN-FS galaxies, while '8311-6104' is the highest optically luminous AGN of the sample. The galaxy '8456-3701' seems to be characterized by standard luminosity and ionization parameter values.

Figure \ref{fig:AGN_discuss} shows $\Delta \alpha$ as a function of L\oiii~$\lambda$5007$_{r < 1 kpc}$ (left panel) and the host's galaxy stellar mass $\log M_*/M_\odot$ (right panel), of the AGN-FS (dark red triangles) and AGN-RPS (red circles) galaxies.
No clear correlations with the stellar mass nor the \oiii~ luminosity are observed when considering the whole sample, either in the AGN-FS or AGN-RPS sample. 
Indeed, for the RPS AGN hosts, the Kendall $\tau$ test shows no correlation in the $\Delta \alpha$ versus $\log$L\oiii$\lambda$5007$_{r < 1 kpc}$ (k$\tau_{AGN} = -0.16$, p-value = 0.36) and $\Delta \alpha$ versus $\log M_*/M_\odot$ (k$\tau_{AGN} = -0.07$, p-value = 0.29) diagrams. 
Similarly for the AGN-FS, according to a Kendall $\tau$ test, the slopes do not correlate with the galaxy's properties ($\tau_{AGN}$ =-0.12, p-value=0.17; and k$\tau_{AGN}$ =0.12, p-value=0.16).
However, the four outliers in Figure \ref{fig:AGN-FSvsRPS} still show steeper slopes in both diagrams, as opposed to JO206 and JO171 which lie in the area covered by the main KDE distribution of the AGN-RPS sample. In other words, the field AGNs have also slopes steeper than galaxies in the same range of nuclear luminosity and host galaxy's stellar mass.

 As an ultimate check, we test if the presence of a stellar bar is influencing the metallicity distributions in our samples, and in particular can explain the peculiar behavior observed in the 6 outliers discussed in this section.
 The influence of the stellar bar on the stellar
and gas components is well-known \citep{berentzen+2007, combes_elmegreen1993}. Previous works showed that the bar appears to flatten the metallicity gradients inside the bar region \citep{zurita+2021a,zurita+2021b} or drive a steeper metallicity gradient \citep{chen+2023}. 
To look for systematic differences induced by the bar in our AGN samples, we divide the galaxies part of the AGN-RPS and AGN-FS into barred and unbarred.
Among the AGN-RPS galaxies, \cite{bacchini+2023} found hints of a stellar bar in JO206, as also in JO201, JO204, and JW100 by 3D modeling the CO gas kinematics observed by ALMA at a resolution of $\sim$ 1 kpc. For the remaining galaxies, we use the classification drawn from the \cite{sanchez-garcia+2023}'s catalog, in which the unbarred galaxies were JO171 and JW39. For the AGN-FS, we use the classification from the  `MaNGA Morphology Deep Learning DR17' catalog \citep{sanchez+2018}. 
The classification of the 'peculiar galaxies' described in this section is of strongly barred for 8311-6104 and 9027-12704 ($P_{\rm bar}$ $\geq 0.98$), while 8993-12705 is at odds with a $P_{\rm bar}$ = 0.48 and 8456-3701 has no bar signature ($P_{\rm bar}$ $\sim$ 0.08). 
In conclusion, we do not report a correlation between the presence of the stellar bar and the peculiar metallicity distributions in these objects.

\begin{figure*}[h]
    \centering
    \includegraphics[width=\linewidth]{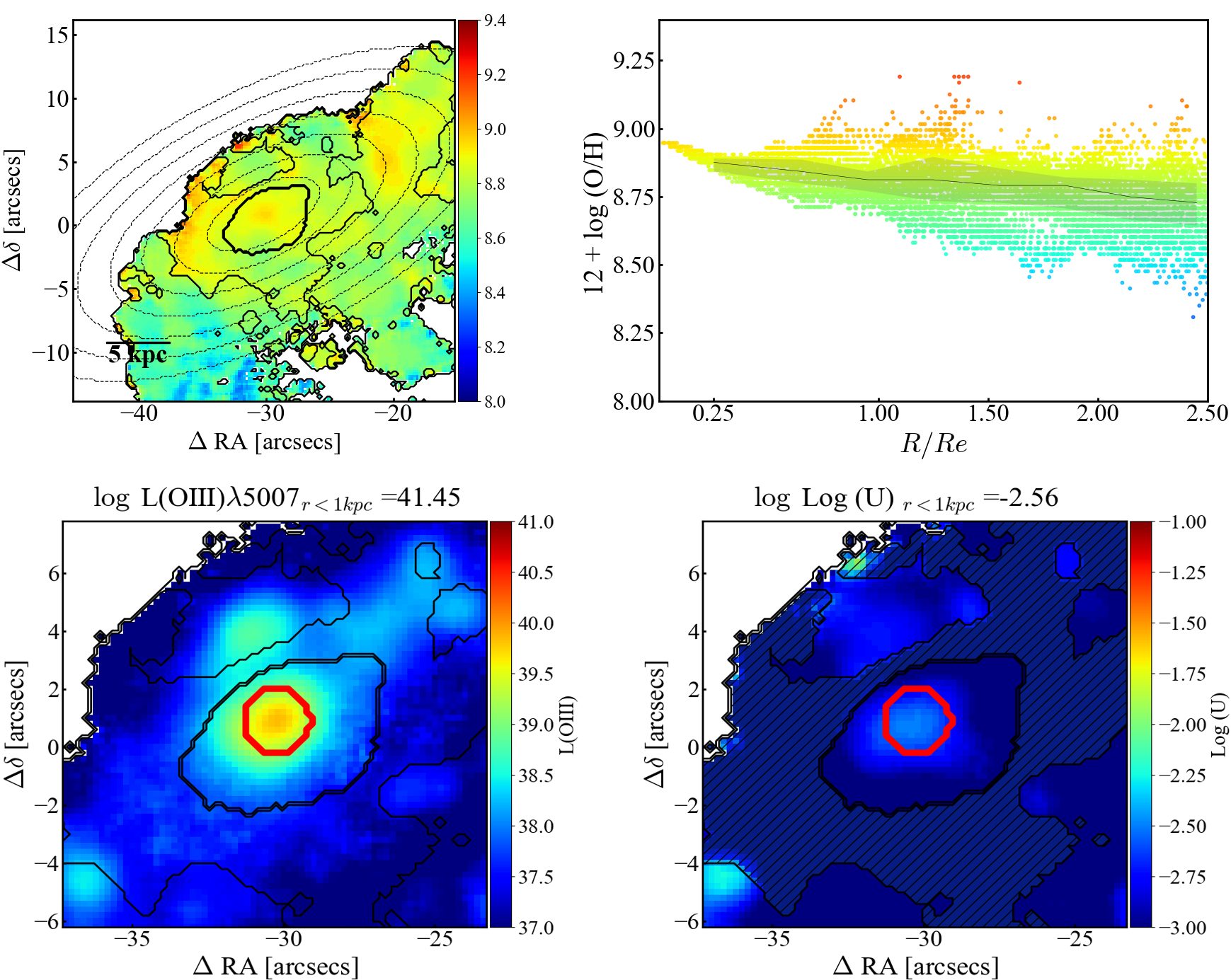}
    \caption{(\emph{Top panels}) Metallicity map (on the left) and metallicity gradient (on the right) of the galaxy 'JO206' ($z=0.051$, $\log M_*/M_\odot$ = 10.96). (\emph{Bottom panels}) Spatially-resolved map of the luminosity of the \oiii$\lambda$5007 line (on the left) and ionization parameter $\log (U)$ (on the right). The red circle encloses the aperture with a (non-de projected) radius of $\sim$ 1 kpc, inside which are computed the median values of $\log L$\oiii~ and $\log U$, also written on top of the plots (e.g. $\log L$\oiii$_{r < 1 kpc}$ and $\log U_{r < 1 kpc}$).}
    \label{fig:pecgal0}
\end{figure*}

\begin{figure*}[h]
    \ContinuedFloat
    \centering
    \includegraphics[width=\linewidth]{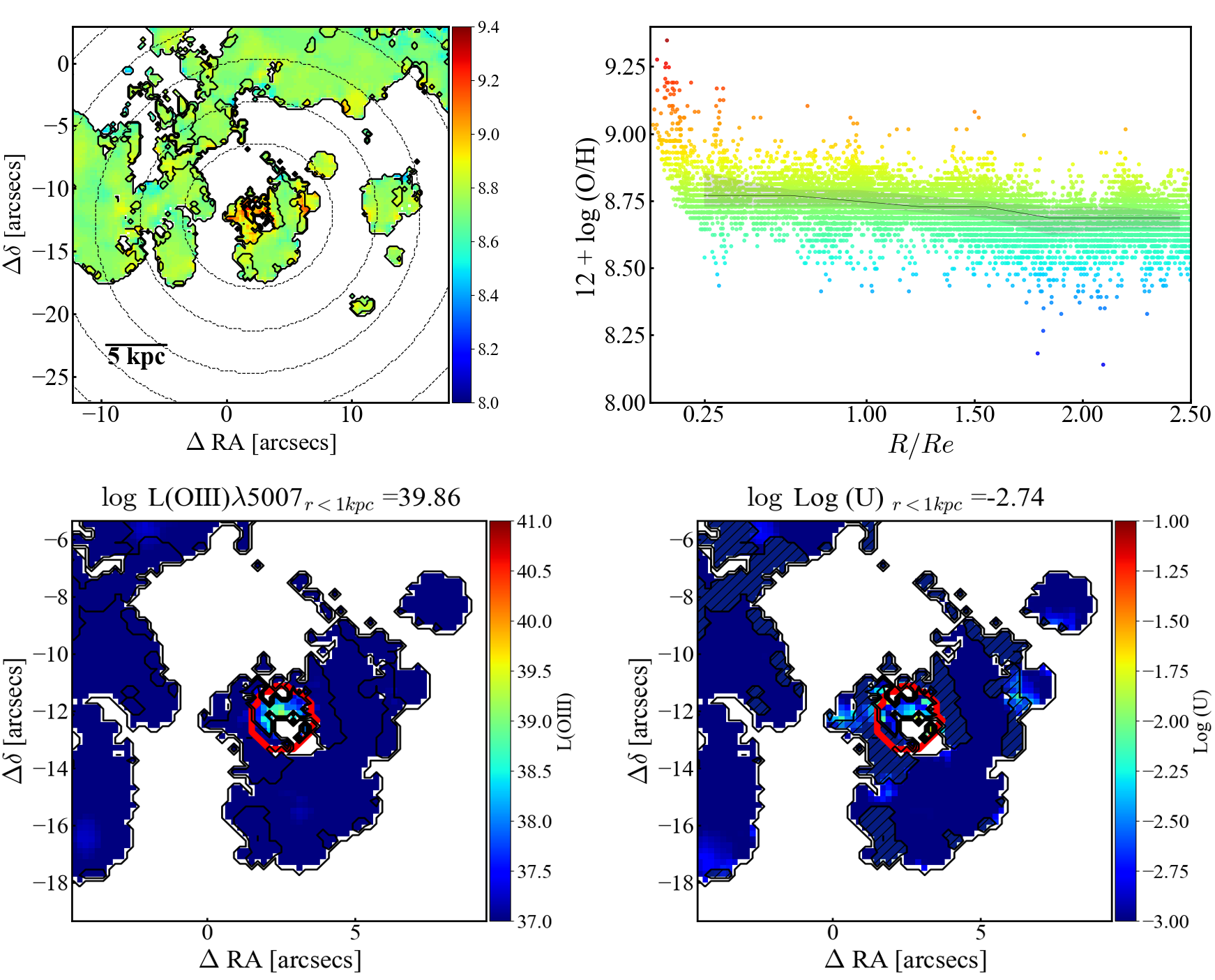}
    \caption{{\it (continued)} Maps and metallicity gradient of the galaxy JO171 ($z=0.052$, $\log M_*/M_\odot$ = 10.61).}
\end{figure*}

\begin{figure*}[h]
    \ContinuedFloat
    \centering
    \includegraphics[width=\linewidth]{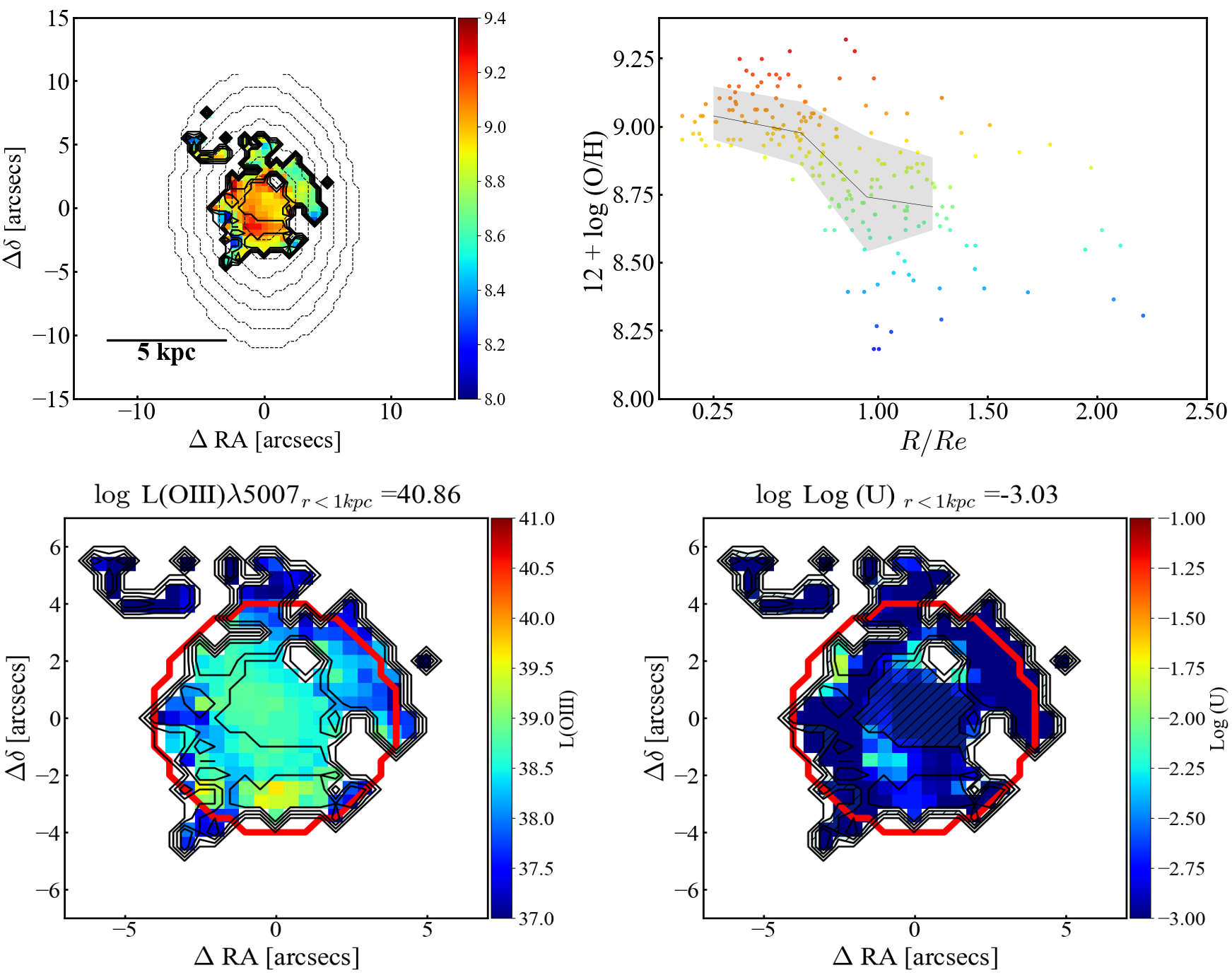}
    \caption{{\it (continued)} Maps and metallicity gradient of the galaxy 8993-12705 ($z=0.029$, $\log M_*/M_\odot = 10.96$). }
\end{figure*}

\begin{figure*}[h]
    \ContinuedFloat
    \centering
    \includegraphics[width=\linewidth]{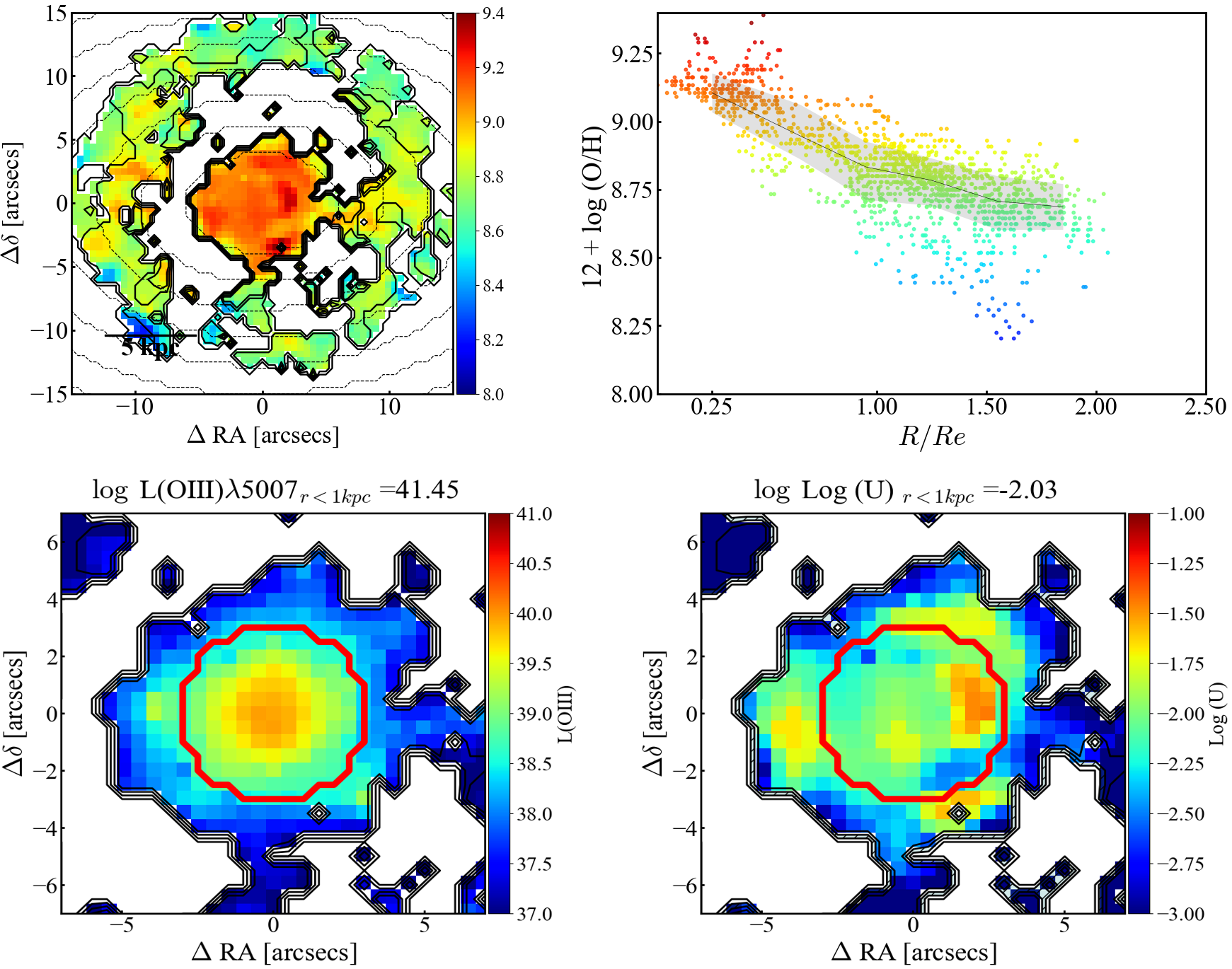}
    \caption{{\it (continued)} Maps and metallicity gradient of the galaxy 9027-12704 ($z = 0.035$, $\log M_*/M_\odot = 10.69$).
    }
\end{figure*}

\begin{figure*}[h]
    \ContinuedFloat
    \centering
    \includegraphics[width=\linewidth]{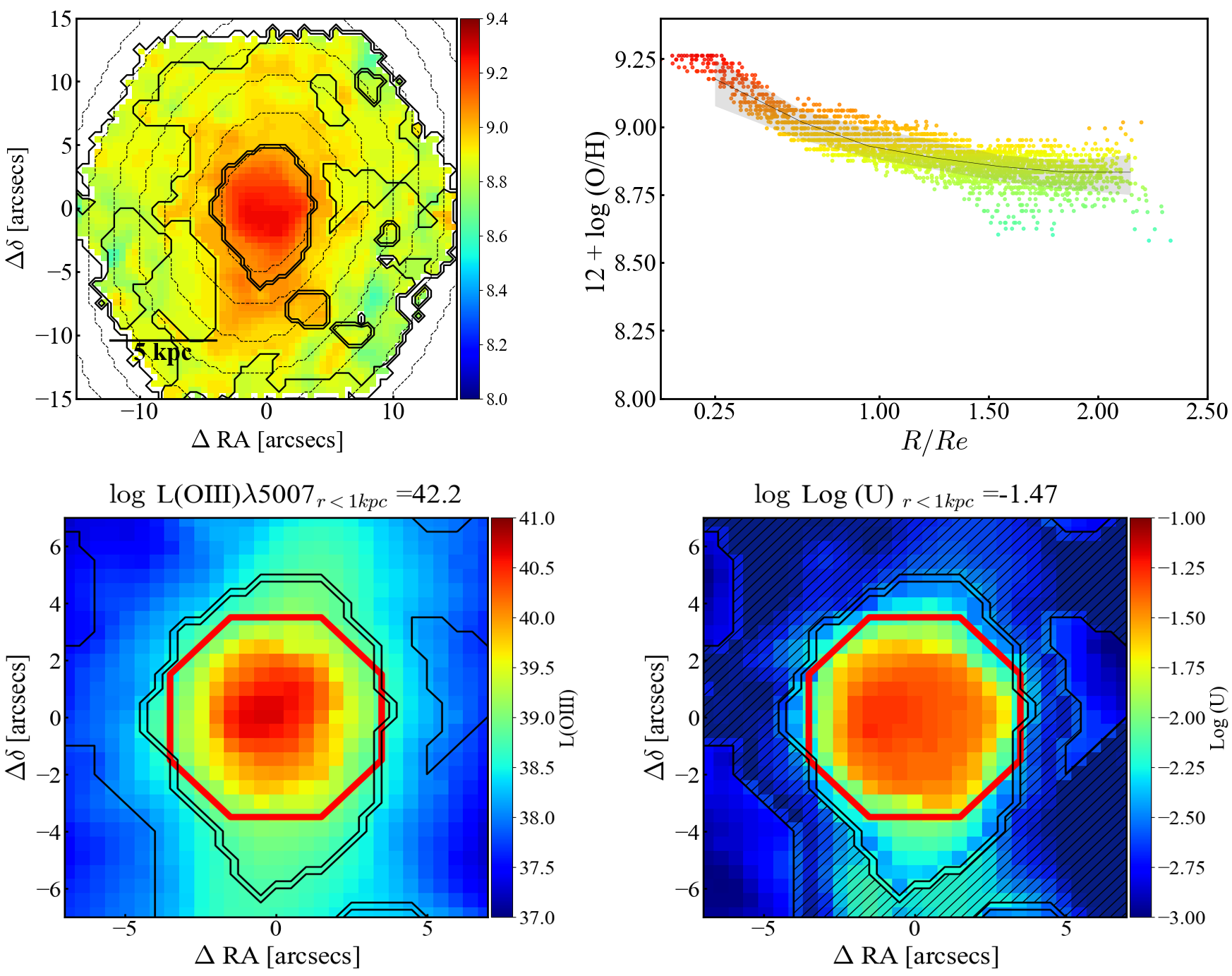}
     \caption{{\it (continued)} Maps and metallicity gradient of the galaxy 8456-3701 ($z = 0.026$, $\log M_*/M_\odot =$ 10.75). }
\end{figure*}

\begin{figure*}[bh]
    \ContinuedFloat
    \centering
    \includegraphics[width=0.9\linewidth]{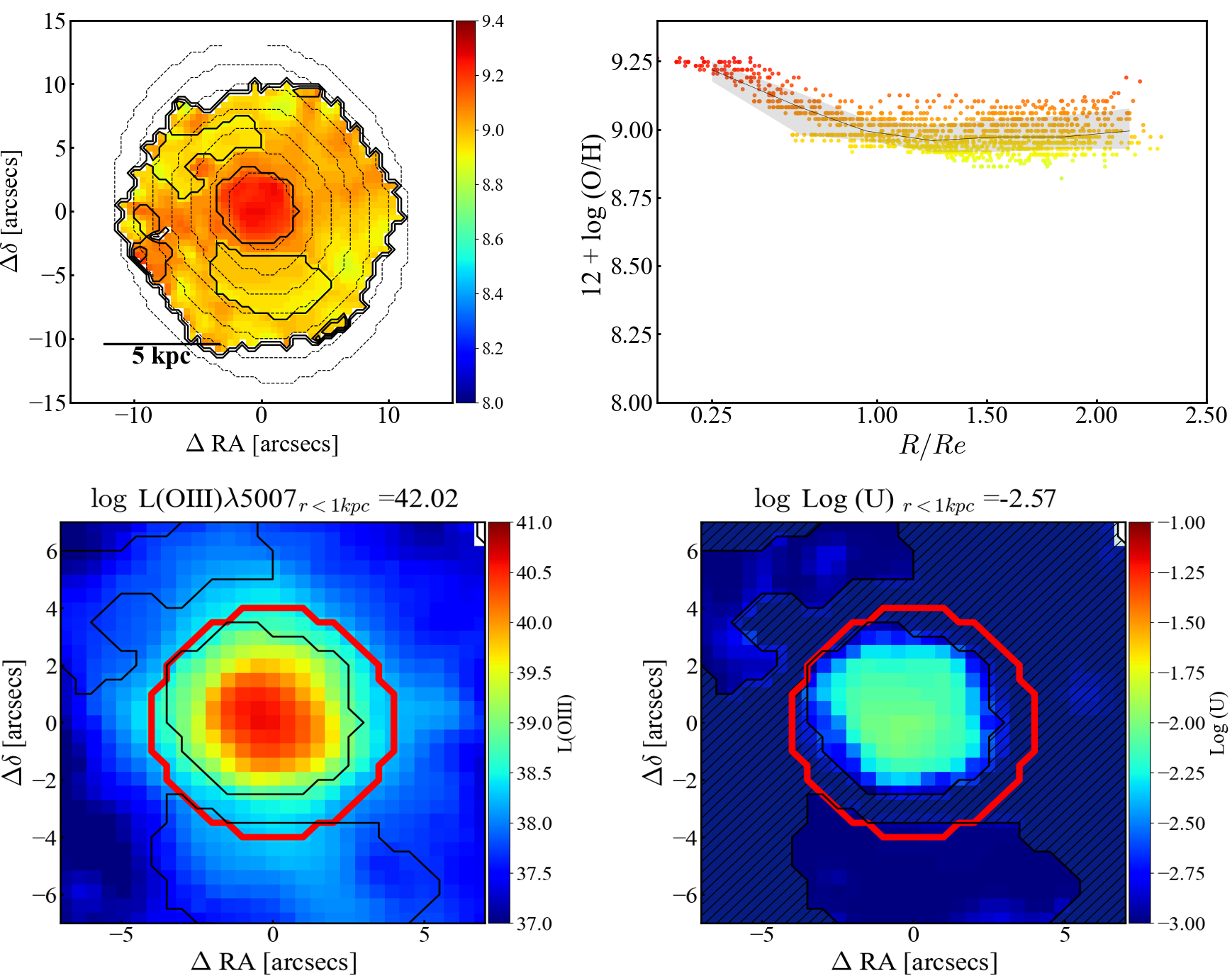}
         \caption{{\it (continued)} Maps and metallicity gradient of the galaxy 8311-6104 ($z=0.027$, $\log M_*/M_\odot = 10.67$).}
\end{figure*}

\begin{figure*}
    \makebox[\textwidth]{
    \includegraphics[scale=0.5]{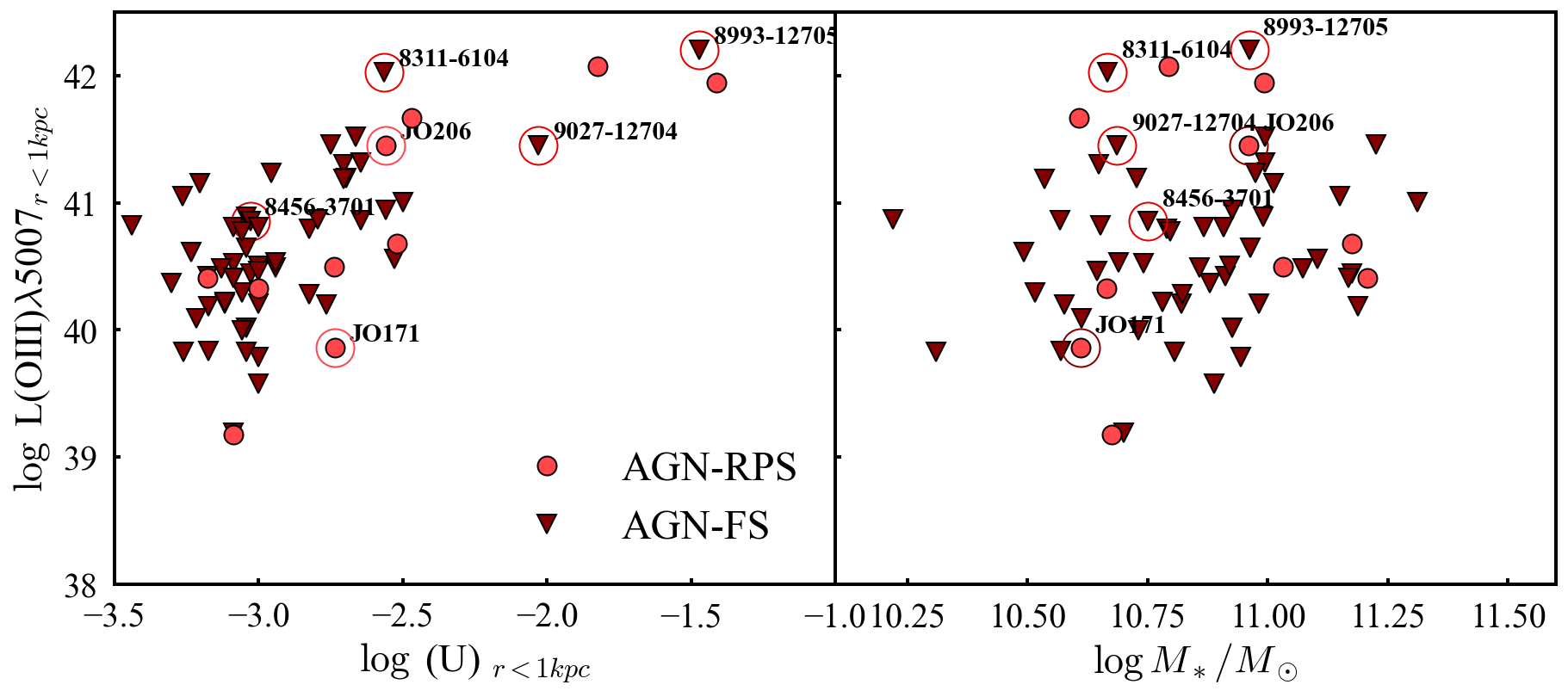}}%
    \caption{Median luminosity of the \oiii$\lambda$5007 line within $r <$ 1 kpc versus the median value of the ionization parameter inside the same aperture (left panel) and versus the host galaxy's stellar mass (panel on the right) of the AGN-FS (dark-red trianges) and AGN-RPS (light-red circles). The 'peculiar' galaxies are highlighted. Overall, the galaxies from the field sample 8311-6104 and 8993-12705 are the most luminous AGNs, with '8993-12705' being also the one with the highest ionization parameter followed by 9027-12704. The AGN hosts JO206, JO171, and '8456-3701' show standard values of both parameters.}
    \label{fig:logu-vs-logo3}
\end{figure*}

\begin{figure*}[ht]
     \centering
     \makebox[0.4\textwidth]{
    \includegraphics[scale=0.6]{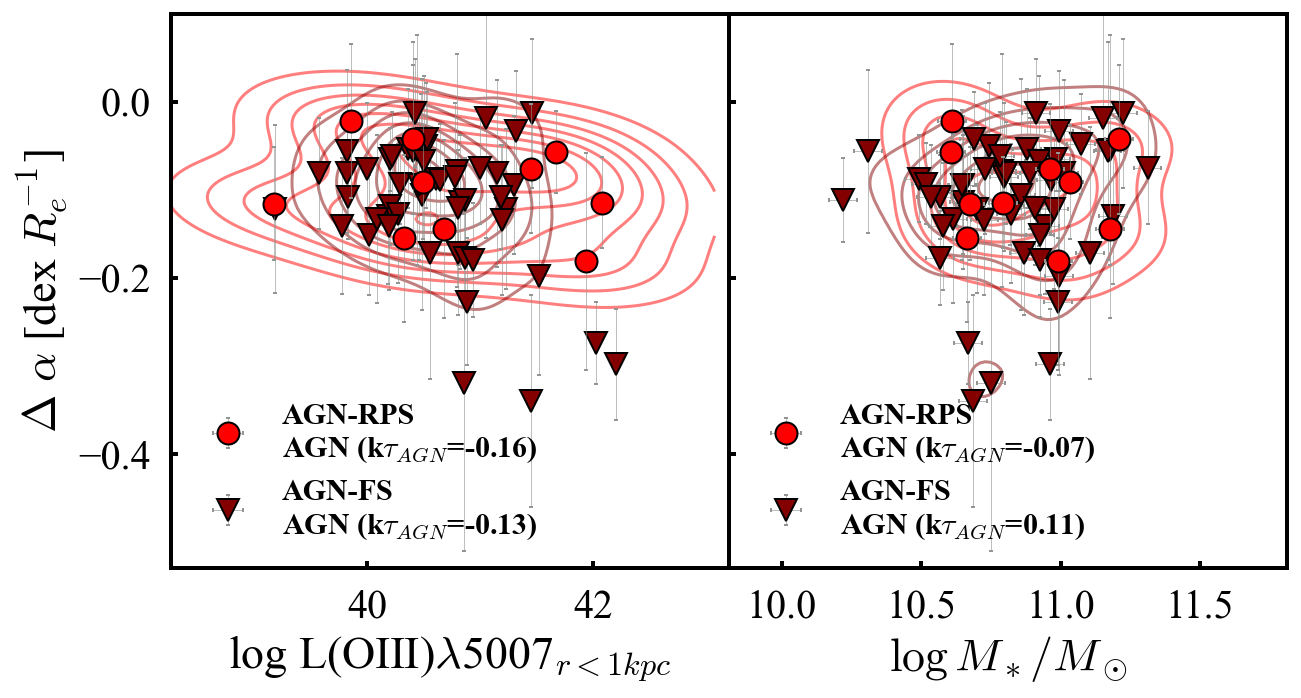}}%
    \caption{Gradient's slope ($\Delta \alpha$) of the AGN-FS (dark red triangles) and AGN-RPS (red circles) as a function of the \oiii~$\lambda$5007 luminosity within an aperture of $r \sim 1$ kpc ($\log$ L \oiii$\lambda$5007$_{r < 1 kpc}$) and of the host's galaxy stellar mass ($\log M_*/M_\odot$). The legend shows the Kendall $\tau$ test's coefficients (e.g. kt$_{AGN}$) for the 10 galaxies part of the AGN-RPS and for the 52 galaxies part of the AGN-FS, revealing that the slopes do not correlate with the other galaxy's properties in both the AGN-RPS and AGN-FS.}
    \label{fig:AGN_discuss}
\end{figure*}

\section{Relation between slopes and stellar mass in star-forming galaxies} \label{sec:appendix-b}

\begin{figure*}
    \makebox[\textwidth]{
    \includegraphics[width=0.7\linewidth]{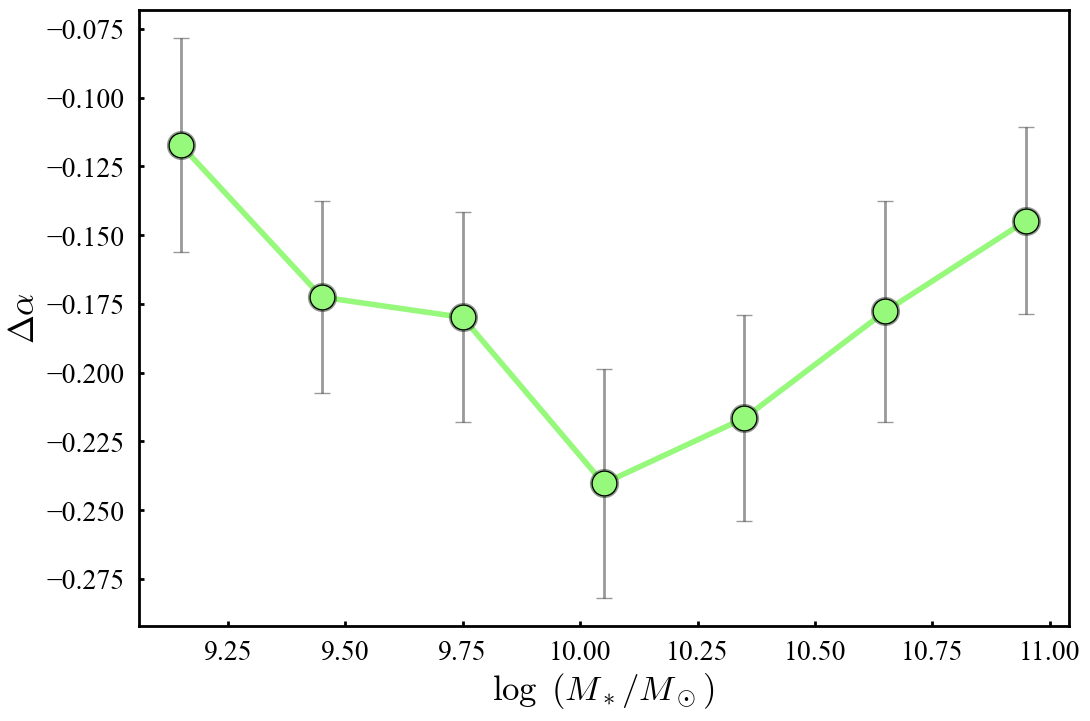}}%
    \caption{Slopes of the SF gradients as a function of the stellar mass. For a comparison to other works, see the shape of the relation derived by \cite{mingozzi+2020} using the code IZI \citep{blanc+2015} coupled with {\sc Mappings IV} photoionization models (e.g. red line in the right panel from Figure 12 of the paper) and by \cite{franchetto+2020} using the PYQZ code \citep{dopita+2013} coupled with the same models as in \cite{mingozzi+2020}.
    }
    \label{fig:sf_slopes}
\end{figure*}

In Figure \ref{fig:sf_slopes}, we show the gradient's slopes of the SF galaxies in both the RPS and field-sample, in the stellar mass range 9.0 $\leq \log M_*/M_\odot \leq $ 11.1.
We observe that the slopes are negative at every stellar mass, ranging between [-0.12, -0.24] dex$R_e^{-1}$. In agreement with the findings of \cite{belfiore+2017b} and \cite{mingozzi+2020} in the MaNGA survey and with the results from \cite{poetrodjojo+2018} in the SAMI survey, we observe a dependence of the SF metallicity gradients on the stellar mass, with the metallicity gradient nearly flat for low-mass galaxies ($M_* \sim 10^9 M_\odot$) and progressively steeper (more negative) for more massive galaxies. We also observe a flattening at stellar mass $M_* \geq 10^{10.3} M_\odot$, similarly to Figure \ref{fig:AGNvsSF_discuss} where the cut in mass is $M_* \geq 10^{10.5} M_\odot$. We attribute the flattening to the saturation of the metallicity at around 12 + log O/H $\sim$ 9.2 typically observed in {\sc Hii} regions, which can readily be explained by classical inside-out chemical evolution models \citep{belfiore+2019b}.

\bibliography{biblio}{}
\bibliographystyle{aa}

\end{document}